\documentclass[twocolumn]{aastex63}

\pdfoutput=1 
\usepackage{amsmath,amstext}
\usepackage[T1]{fontenc}
\usepackage{gensymb}
\usepackage{subfigure}
\usepackage{longtable}
\usepackage[inline]{enumitem}
\usepackage{cleveref}
\crefname{figure}{Figure}{Figures}
\crefname{table}{Table}{Tables}
\crefname{equation}{equation}{equations}
\usepackage{soul}
\usepackage{ulem}
\usepackage{savesym}
\savesymbol{tablenum}
\usepackage{siunitx}
\restoresymbol{SIX}{tablenum}
\DeclareSIUnit{\year}{yr}
\DeclareSIUnit{\mas}{mas}
\DeclareSIUnit{\astronomicalunit}{au}
\DeclareSIUnit{\parsec}{pc}
\usepackage{xcolor}

\newcommand{\blue}[1]{\textcolor{blue}{#1}}
\newcommand{\redsout}{\bgroup\markoverwith{\textcolor{red}{\rule[0.5ex]{2pt}{0.4pt}}}\ULon}

\submitjournal{ApJS}
\accepted{February 19, 2020}

\shorttitle{The SUPERWIDE Catalog}
\shortauthors{Hartman $\&$ L\'{e}pine}
\graphicspath{{./}{figures/}}

\begin{document}

\title{The SUPERWIDE Catalog: A Catalog of \num{99203} Wide Binaries Found in Gaia and Supplemented by the SUPERBLINK High Proper Motion Catalog}

\author[0000-0003-4236-6927]{Zachary D. Hartman}
\affiliation{Lowell Observatory, 1400 W Mars Hill Road, Flagstaff, AZ 86001}
\affiliation{Georgia State University,Department of Physics \& Astronomy, Georgia State University, 25 Park Place South, Suite 605, Atlanta GA, 30303, USA}

\author[0000-0002-2437-2947]{S\'{e}bastien L\'{e}pine}
\affiliation{Georgia State University,Department of Physics \& Astronomy, Georgia State University, 25 Park Place South, Suite 605, Atlanta GA, 30303, USA}

\nocollaboration{2}



\begin{abstract}

We present a catalog of \num{99203} wide binary systems, initially identified as common proper motion (CPM) pairs from a subset of \SI{\sim 5.2}{million} stars with proper motions $\mu > \SI{40}{\mas\per\year}$, selected from Gaia data release 2 (DR2) and the SUPERBLINK high proper motion catalog. CPM pairs are found by searching for pairs of stars with angular separations \SI{< 1}{\degree} and proper motion differences $\Delta\mu < \SI{40}{\mas\per\year}$. A Bayesian analysis is then applied in two steps. In a first pass, we use proper motion differences and angular separations to distinguish between real binaries and chance alignments. In a second pass, we use parallax data from Gaia DR2 to refine our Bayesian probability estimates. We present a table of 119,390 pairs which went through the full analysis, 99,203 of which have probabilities $>95\%$ of being real wide binaries. Of those 99,203 high probability pairs, we estimate that only about 364 pairs are most likely to be false positives.  In addition, we identify 57,506 pairs which have probabilities greater than $10\%$ from the first pass but have high parallax errors and therefore were not vetted in the second pass. We examine the projected physical separation distribution of our highest probability pairs and note that the distribution is a simple exponential tail, and shows no evidence of being bi-modal.  Among pairs with lower probability, wide binaries are detected at larger separations ($>10^{4-5}$ AU) consistent with the very wide population suggested in previous studies, however our analysis suggests that these do not represent a distinct population, but instead represent either the exponential tail of the ``normal'' wide binary distribution, or are simply chance alignments of unrelated field stars. We examine the HR diagram of this set of high probability wide binaries and find evidence for 980 over-luminous components among 2,227 K+K wide binaries; assuming these represent unresolved sub-systems, we determine that the higher order multiplicity fraction for K+K wide systems is at least 39.6\%.

\end{abstract}



\section{Introduction}
\label{sec:intro}
Binary star systems have long been important tools for stellar astrophysics, as they can be used to determine the physical properties of their stars (e.g. their masses and radii), which can be difficult for single stars.  They are also relatively common: roughly half of the solar type stars in the local neighborhood are found to be in binary systems \citep{1991duquennoy,2010Raghavan}.  Low-mass stars, such as the ubiquitous M dwarfs, have been found to have a multiplicity fraction of $26.8 \pm 1.4\%$ \citep{2019wintersmdwarfmult}.  They can range from close spectroscopic/eclipsing binaries with separations on the order of the Sun's radius to wide visual binary systems with separations that can reach \SI[retain-unity-mantissa=false]{> e4}{\astronomicalunit}.  

These wide visual binary systems represent important laboratories for stellar astrophysics.  The two components in wide binary systems are too far apart to be interacting in any meaningful way at the present time, and are thus equivalent to two single stars evolving in unison.  Monitoring the motion of binary stars makes it possible to independently measure their gravitational masses without having to rely on a stellar model.  In addition, because the two components are thought to form coevally from adjacent parts of a molecular cloud, these systems are crucial for calibrating gyrochronology relations \citep{2012chanamegyro,2017Janes,2018Godoy} and metallicity scales \citep{2007lepinemetal,2013Mannmetal,2014Newtonmetal,2014Mannmetal,2017Veyettemetal,2018AndrewsChanamemetal,2019Andrewsapoggee}, in particular to determine how age and metallicity diagnostics vary with stellar mass.  Wide binary stars are also excellent tools for examining phenomena that evolve over time, such as stellar activity and flaring behavior \citep{2014Gunningflare,2016Morganflare,2018WideFlareKepler}.  

Another important test that wide binary systems can provide is in the area of Galactic dynamics. As the binding energies of wide binary systems should be small, they are easily disrupted by interactions with other stars or with Galactic tidal fields.  This allows one to set limits about the stellar density, mass functions, and general environment of various Galactic locales \citep{1987weinberg,2009Parker,2010jiang}.  We can also place limits on the number density and typical masses of some dark matter candidates such as MACHOs \citep{2004ChanameGould}, which are expected to act as gravitational disrupters of wide binary systems.  

Wide binaries are also important for exoplanet research, since many planets have been found orbiting stars that also have wide stellar companions \citep{2016Deacon}.  One might expect a companion, even a relatively distant one, to have potentially disruptive effects on the structure of the protoplanetary disk, which may affect the formation of planets.  In particular, orbits for these planets are expected to be mildly to highly eccentric.  An expansion of the work of \citet{2016Deacon} beyond Kepler to compare the frequency of exoplanet systems in wide binaries versus single stars may shed light on planet formation mechanisms. 

A more fundamental question regarding the widest binaries ($\rho > \SI{10000}{\astronomicalunit}$) is how they form in the first place.  These systems can have separations larger than a typical protostellar core (\SI{\sim 0.1}{\parsec}; \citealp{2010Kouwenhoven,2013Duchene,2012Reipurth,2017Tokovinin}).  There have been several proposed channels for how these systems form, including the cluster dissolution scenario \citep{2010Kouwenhoven}, the unfolding of higher order multiples \citep{2012Reipurth}, and the adjacent cores scenario \citep{2017Tokovinin}.  Recent evidence from young moving groups and star forming regions \citep{2016Elliott, 2017Joncour} have found that these wide binaries are found mostly as part of higher order multiples.  However, more work is needed to determine which, if any, mechanism is the dominant formation channel for the widest wide binaries.  In particular, determining whether there is a difference between Disk and Halo wide binaries could yield interesting results.

Previous searches for wide binaries have focused on finding pairs of stars that are close to each other on the sky and have similar proper motions and/or similar estimated distances \citep{2004ChanameGould,2007lepine,2010Dhital,2011Lepine,2011Shaya,2012Toko,2014Toko,2015slowII,2016Deacon,2017AndrewsChaname,2017Oh,2017Oelkers,2018elbadry,2018coronadohalowbs,2019Jimenez}.  Pairs of gravitationally bound stars with very large orbital separations (\SI{> 10000} AU) have orbital periods that are long enough (\SI{> 700000}{years}) that the orbital velocity should be very low (\SI{\geq 0.3}{\kilo\metre\per\second}).  As a result, the contribution of the orbital motion to the total space motion should be negligible in most cases, and both components will have near-identical proper motions.  One can thus identify wide binaries by looking for close pairs of stars with similar proper motions; these are typically called common proper motion (CPM) pairs.  However, coincident alignment, compounded by measurement uncertainties, can make two unrelated stars appear as a CPM pair by chance. In this case, additional work must be performed to confirm the pair is gravitationally bound.  This can be done with a variety of methods, from obtaining more precise proper motion measurements to measuring radial velocities.  Ultimately, one would want to confirm the spatial proximity of the two stars using accurate parallax data.

With the advent of Gaia, this field of astronomy has entered a new phase.  Gaia Data Release 1 (DR1; \citealp{2016gaiadr1a,2016gaiadr1b}) has already provided accurate proper motions, positions and distances for stars in the Tycho-Gaia astrometric solution (TGAS) \citep{2015Michalikgaia}.  Three groups conducted separate searches of the TGAS catalog for wide binaries.  \citet{2017Oelkers} examined TGAS and combined it with the Sloan Digital Sky Survey (SDSS) to produce their catalog of \num{8660} possible wide pairs.  This catalog was a mix of TGAS-TGAS and TGAS-SDSS pairs, and expanded the work of \citet{2015slowII} to higher mass stars.  Their method made use of a Galactic model similar to the one used in \citet{2015slowII} to calculate the probability that a given pair is a chance alignment based on predictions of the local field density.  

\citet{2017Oh}, on the other hand, searched for wide binaries by calculating a likelihood ratio that a pair in TGAS is a co-moving pair based on its tangential velocity and physical separation.  Their search not only found wide binaries but also co-moving groups such as open clusters and OB associations.  They found \num{13058} high probability candidate pairs in their search.  Both \citet{2017Oh} and \citet{2017Oelkers} claim to have found evidence that there was an excess of pairs at wide separations and pairs with separations greater than \SI{1}{\parsec}.  Both argue that this excess of pairs consists of a population of wide stars that are either very loosely bound or not bound at all and are just the remains of wide binaries that have been torn apart.  Both also point out that this population of pairs at large separations should be relatively young, as these unbound pairs would have drifted much father apart were they older than a few hundred million years. \citet{2017Oelkers} goes further and reinforces the proposal by \citet{2010Dhital} that this excess is the result of a second formation scenario from the dissolution of clusters of stars.  

The third search was conducted by \citet{2017AndrewsChaname}.  They ran a Bayesian analysis of the TGAS catalog taking into account angular separations, proper motion differences, and parallax differences.  After removing known open clusters, they identified \num{7108} candidate wide binary pairs.  They also matched their catalog to the RAdial Velocity Experiment (RAVE; \citealp{2017RAVE}) survey and found a number of their pairs to have radial velocities (RVs) in the survey.  They compared RVs and found that the majority of their pairs had similar RVs, confirming that they are binaries.  However, they also found that for pairs with projected physical separations larger than \SI{4e4}{\astronomicalunit}, only about half of the pairs had matching RVs, which was confirmed in their follow-up paper \citep{2018Andrewsrv}.  This appears to lend support to the argument that pairs with separations greater than \SI{\sim 1}{\parsec} may not be genuine wide binaries but simply chance alignments of unrelated field stars.

More recently, Gaia Data Release 2 (DR2; \blue{\citealp{2018Gaiadr2,2018Lindegren}}) has expanded the possible search area for wide binaries.  One search has already been conducted on this new catalog.  \citet{2018elbadry} examined the ``cleaned'' Gaia DR2 catalog (\citealp[see][]{2018Lindegren} for more details) for wide binaries using two cuts.  The first was a cut in actual physical separation of the pairs set at \SI{50000} AU while the other was a cut in proper motion space which depended on several parameters including distance and angular separation of the pairs.  The result of this cut and the removal of clusters, moving groups and resolved higher order multiples is a ``pure'' sample of \num{\sim 53400} wide binaries, although there still exists unresolved higher order multiples, as noted by the authors.  They also claimed to see a difference in the distribution of physical separations between three types of wide binaries -- main sequence + main sequence, white dwarf + main sequence and white dwarf + white dwarf -- which they claim to be caused by a kick during the white dwarf formation \citep[see][sections ~3-4]{2018elbadry}.  In section 5.1.2 of this paper, we compare our own sample of binaries to that of \citet{2018elbadry} and find strong agreement between the two catalogs.  

In this paper, we develop a Bayesian approach to conduct our own common proper motion search for wide binaries in the high proper motion subset of Gaia DR2 stars with proper motions greater than \SI{40}{\mas\per\year}.  We define a wide binary in the same way as \citet{2017AndrewsChaname}, namely any high probability pair we identify we consider to be a wide binary based on the data we have available to us.  The Gaia DR2 high proper motion subset contains \SI{\sim 5.2}{million} high proper motion stars; our search ultimately identifies \num{99106} wide binary candidates with \SI{> 95}{\percent} probability of being gravitationally bound systems.  

The outline of the paper is as follows: Section~\ref{sec:initialSearch} briefly describes the initial search catalog.  We then elaborate on the method of \citet{2007lepine} for artificially constructing a randomized set of stars completely devoid of wide binaries, which we use as a reference sample to estimate the occurrence of chance alignments; this is explained in Section~\ref{sec:wideID}, along with our Bayesian method to estimate the probability of a pair being a true binary based on positions and proper motions only.  In Section~4, we expand our analysis to incorporate parallax data from the Gaia DR2.  Section~5 shows several checks on our final result, including radial velocity confirmation. In Section~6, we perform an analysis of the resulting catalog of wide systems.  We summarize our conclusions in Section 7.

\section{Initial Search Catalog}
\label{sec:initialSearch}
With over \SI{1.3}{billion} sources with positions, proper motions and parallaxes, the Gaia Data Release 2 (DR2) catalog offers an excellent data set to search for wide binaries. \citep{2016Gaiaa,2016Gaiab,2016Lindegren,2018Gaiadr2}.  Average parallax errors range from \SI{0.04}{\mas} for bright targets ($G < \SI{15}{mag}$) to \SI{0.7}{\mas} for the faintest targets at $G = \SI{20}{mag}$.  For that same range of magnitudes, the average proper motion errors are \SI{0.06} to \SI{1.2}{\mas\per\year}.  In order to reduce the impact of potential chance alignments, i.e. stars that are close to each other on the sky but not related, we only considered stars with proper motions larger than \SI{40}{\mas\per\year}.  As will be shown in Section 3, pairs that have higher proper motions are more likely to be real binaries because there are fewer stars with high proper motions.  Instead of having to deal with \SI{\sim 1}{billion} stars, this proper motion cut leaves a more managable subset of about \SI{5.2}{million} sources. 

We supplemented this Gaia subset with stars from the SUPERBLINK high proper motion catalog \citep{2005lepine,2011lepinegaidos}. This catalog lists \SI{2.7}{million} stars with proper motions \SI{> 40}{\mas\per\year}.  In addition to limiting the number of chance alignments, our proper motion cut allows our sample from Gaia DR2 to match with that of SUPERBLINK. It is an all-sky catalog complete to a proper motion limit of \SI{40}{\mas\per\year} for declinations from \SI{+90}{\degree} to \SI{-30}{\degree} and \SI{80}{\mas\per\year} for declinations south of \SI{-30}{\degree}.  It was updated with the Gaia first data release \citep{2016Gaiaa,2016Gaiab,2016Lindegren}, incorporating the more accurate Gaia proper motions from DR1 at the brighter end ($V<12$) and combining the 2016-epoch Gaia DR1 positions with the 2000-epoch positions of the stars in the 2-Micron All-Sky Survey (2MASS) catalog to obtain more accurate proper motions at the fainter end ($12<V<20$), \citet{2006AJ2mass}.  The nominal accuracy of the proper motions in the SUPERBLINK-GAIA DR1 catalog are estimated to be \SI{\pm 4}{\mas\per\year}.  However, the proper motion accuracy is higher at the brighter end because the more accurate proper motion values from the TGAS catalog are used.  

One advantage of using the SUPERBLINK catalog is that all the stars have been verified using various quality control checks, which include visual inspection of Palomar Sky Survey images using a blink comparator-type software.  Common proper motion pairs in particular have been extensively targeted for visual inspections, and the rate of false identifications is expected to be less than \SI{0.1}{\percent}.  In many cases, pairs were identified that appear single on Palomar Sky Survey images, but that are clearly resolved as close pairs on 2MASS images - the higher proper motion is normally sufficient to rule out chance alignments with background field stars by comparing the 1999-2000 images from 2MASS with the 1950s images from the Palomar Sky Survey.  While most of the stars in SUPERBLINK are in the Gaia DR2 catalog, there are about \num{64000} stars in SUPERBLINK that are not in Gaia DR2.  This was after a match was attempted using a position matching algorithm that took proper motion into account and made the final match to the star with the comparable magnitude to the star in question. These missing DR2 stars notably include some with very large proper motions, and also likely include stars with irregular astrometric solutions such as nearby astrometric binaries.  Recent work by \citet{2018roboaogaiacheck} has shown that Gaia does not systematically include binaries with angular separations between 0-2 arcseconds; many such pairs are however properly recorded in the SUPERBLINK catalog.  All these deficiencies make the Gaia catalog somewhat biased against nearby visual and astrometric binaries and thus also biased against nearby wide systems with a tertiary component; using the SUPERBLINK catalog mitigates some of these biases. 

The original SUPERBLINK catalog used a proper motion lower limit of \SI{40}{\mas\per\year} and had a nominal proper motion accuracy of \SI{\pm 8}{\mas\per\year}.  With the revised proper motion measurements obtained from the inclusion of positional data from Gaia DR1, some stars are found to have proper motions below that limit.  These stars were kept in the catalog, which means that at the present time, the proper motion limit of the catalog does not cut sharply at \SI{40}{\mas\per\year} but has a smooth edge around that limit.  In a small number of cases (0.8\% of the catalog), some stars in the original SUPERBLINK were found to have significantly smaller proper motions ($\mu < \SI{20}{\mas\per\year}$) after the Gaia correction.  We now believe these stars to be ``false positives``, i.e., stars that were incorrectly identified as high proper motion stars in the original SUPERBLINK analysis.  These stars failed the quality controls for a number of reasons, and tend to be concentrated in areas of the southern sky where proper motion uncertainties are significantly higher.  For this analysis, however, we exclude any star with $\mu < \SI{39.8}{\mas\per\year}$.  In addition, as stars may be present in more than one of the catalogs described above, the order of which proper motion is used for this search is as follows: Gaia DR2, SUPERBLINK+Gaia DR1, SUPERBLINK.  This means that if a star has a Gaia DR2 proper motion, this is the proper motion used, but if a star is not present in Gaia DR2 or DR1, then the SUPERBLINK proper motion is used.

\section{Wide System Identification Method: First Pass Using Proper Motions}
\label{sec:wideID}
\subsection{Bayesian Search for True Binaries: Real Pairs vs. Chance Alignments}
\label{sec:wideIDInit}
Starting with the combined SUPERBLINK+Gaia subset of \SI{5.2}{million} high proper motion stars, we search for all pairs of stars with angular separations \SI{2}{\arcsec} < ($\rho$) < \SI{1}{\degree} on the sky and proper motion difference magnitudes less than \SI{40}{\mas\per\year}.  These relatively wide limits are used to ensure that all wide physical systems would be found, at the expense of also including a large number of chance alignments, to be cleaned later.  The angular separation lower limit of \SI{2}{\arcsec} is a conservative estimate of the effective resolving power of the Gaia survey and is set to ensure that our sample of CPM pairs is complete within that range of angular separations \citep{2018roboaogaiacheck} so that we can model the distribution of angular separations as discussed in Section 3.4.  In addition, we limit the magnitude of the primary star to brighter than 19th magnitude.

Given the relatively large upper search radius of \SI{1}{\degree}, we simplify the mathematical algorithm by converting the right ascension ($\alpha$) and declination ($\delta$) coordinates into unit vectors in a 3D Cartesian system $\begin{pmatrix}x & y & z\end{pmatrix}$ via
\begin{align}
    x &= \cos(\delta) \cos(\alpha) \label{eq:xCoord} \\
    y &= \cos(\delta) \sin(\alpha) \label{eq:yCoord} \\
    z &= \sin(\delta)\; . \label{eq:zCoord}
\end{align}
Likewise, we convert all proper motion $\mu_{\alpha}$ and $\mu_{\delta}$, which are locally vectors in the plane of the sky, into their equivalent vectors in 3D Cartesian space $\begin{pmatrix}\mu_x & \mu_y & \mu_z\end{pmatrix}$ using
\begin{align}
    \mu_x &= \mu_\alpha \sin(\alpha) - \mu_\delta \sin(\delta) \cos(\alpha) \label{eq:muX} \\
    \mu_y &= \mu_\alpha \cos(\alpha) - \mu_\delta \sin(\delta) \sin(\alpha) \label{eq:muY} \\
    \mu_z &= \mu_\delta \cos(\delta)\; . \label{eq:muZ}
\end{align}
This vector method for calculating angular separations minimizes problems near the celestial poles.  We calculate the angular separations between any two stars from the dot product definition of their unit position vectors, while the proper motion difference between the two stars are calculated from the magnitude of the difference between their proper motion vectors.  This initial search yields a list of \num{\sim 557000000} CPM pairs from the combined catalog; the overwhelming majority of these ``pairs'' are of course chance alignments, i.e. stars that happen to be near each other on the sky but are not physically related to each other.

To determine which of the 557 million pairs may be true binary systems, we conduct a Bayesian analysis of the complete list of CPM pairs identified in Section 3.1 using a two-step process.  First, we model the statistical distribution of angular separations and proper motion differences for both physical pairs and chance alignments, without considering of parallax information.  We convert all coordinates and proper motion vectors into the Galactic coordinate system, obtaining $l$, $b$, $\mu_l$ and $\mu_b$ for all stars; the use of the Galactic coordinate system will become clear later.  For each pair, we calculate the proper motion difference in the Galactic longitude and Galactic latitude $\Delta\mu_l$ and $\Delta\mu_b$.  We then examine the statistical distribution of any subset of pairs as a function of $\rho$, $\Delta\mu_l$, and $\Delta\mu_b$.

The Bayesian formula for the probability of any pair to be a physical binary given their proper motion difference and angular separation is
\begin{equation}
\label{eq:pairProb}
P\left(B\mid\Delta\mu_l,\Delta\mu_b,\rho\right)=\frac{P\left(\Delta\mu_l,\Delta\mu_b,\rho\mid B\right)\, P(B)}{P\left(\Delta\mu_l,\Delta\mu_b,\rho\right)} \; ,
\end{equation}
where 
\begin{equation*}
\begin{split}
P\left(\Delta\mu_l,\Delta\mu_b,\rho\right) &= P\left(\Delta\mu_l,\Delta\mu_b,\rho\mid B\right)\, P(B) \\
&+ P\left(\Delta\mu_l,\Delta\mu_b,\rho\mid\bar{B}\right)\,P\left(\bar{B}\right) \; .
\end{split}
\end{equation*}
Here, $B$ represents the hypothesis that a pair is a physical binary, and $\bar{B}$ the hypothesis that it is not a binary, i.e. that it is a chance alignment.  As proper motion difference and angular separation are expected to be independent for real binaries and chance alignments, the probability $P\left(\Delta\mu_l,\Delta\mu_b,\rho\mid B\right)$ from the first term in equation (8) can be split into $P\left(\Delta\mu_l,\Delta\mu_b\mid B\right) P\left(\rho\mid B\right)$ and similarly for $P\left(\Delta\mu_l,\Delta\mu_b,\rho\mid \bar{B}\right)$ from the second term.  However, $\Delta\mu_l$ and $\Delta\mu_b$ are correlated, in particular for chance alignment pairs (as will be shown later), and thus their probability density function cannot be written as the product of two independent probabilities.  Therefore, in order to use \cref{eq:pairProb}, we need to calculate the four different probability density functions, $P\left(\Delta\mu_l,\Delta\mu_b\mid B\right), P\left(\rho\mid B\right)$, $P\left(\Delta\mu_l,\Delta\mu_b\mid \bar{B}\right)$ and $P\left(\rho\mid \bar{B}\right)$, in addition to the two priors, $P(B)$ and $P\left(\bar{B}\right)$, which represent the odds for any CPM pair to be either a real pair or a chance alignment respectively.  

\begin{figure*}
\centering
\subfigure{
\includegraphics[scale=0.36]{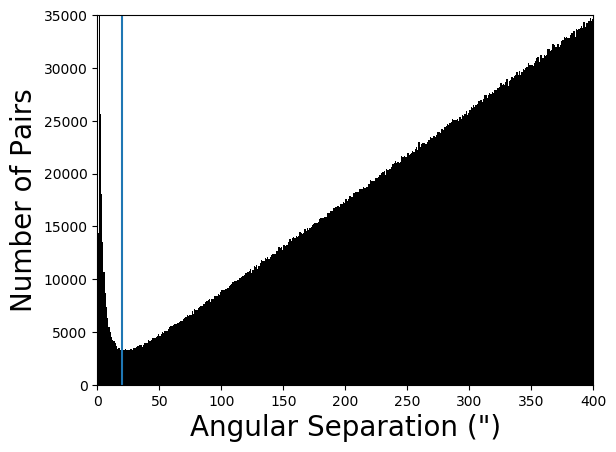}}
\subfigure{
\includegraphics[scale=0.36]{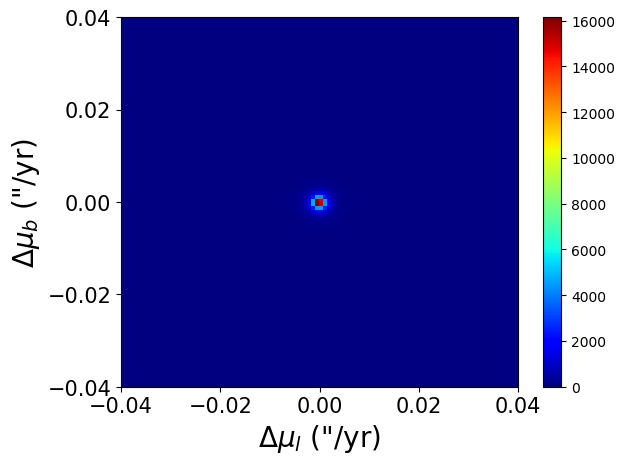}}
\subfigure{
\includegraphics[scale=0.36]{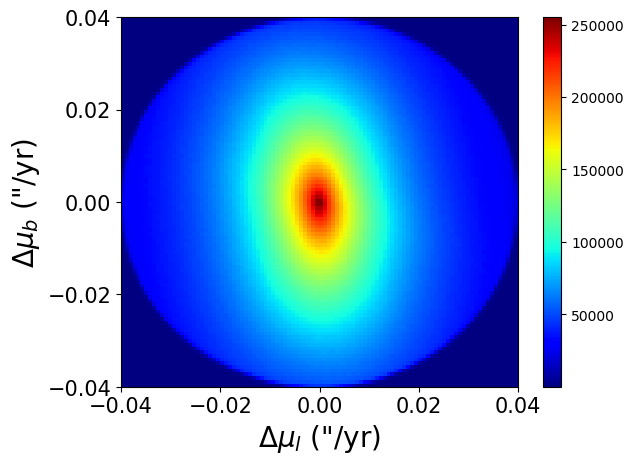}}
\caption{\label{fig:globalreal}Distribution of angular separation and proper motion differences for all pairs found in the initial search of the Gaia catalog subset.  The left panel shows the overall distribution of angular separations.  The middle panel shows the proper motion difference distribution of the close pairs with angular separations less than \SI{20}{\arcsec} (left of the blue line in the histogram), likely representing physical companions.  The right panel shows the proper motion difference distribution of wide pairs with angular separations larger than \SI{20}{\arcsec} (right of the blue line in the histogram), likely representing chance alignments.}
\end{figure*}

The four probability distribution functions described above can be determined empirically (i.e., directly from our initial set of possible CPM pairs) using methods that will be described below.  Our empirical approach stands in contrast
to methods proposed elsewhere, as in \citet{2018elbadry}, who
use a simple cut in physical separation and proper motion difference
to separate physical pairs from chance alignments. Our method also
differs somewhat from the chance alignment estimates of \citet{2015slowII} and \citet{2017AndrewsChaname}, where probability distribution functions are determined from a semi-empirical model of the Galaxy
that describes the stellar density and kinematics using simple exponential/power laws. The advantage of our more direct approach
is that is does not rely on any particular assumption about the
functional form of the local density/kinematics of field stars,
and may thus better account for local fluctuations or substructure
in the spatial or velocity-space distribution of nearby stars. To figure out the probability distribution functions of the physical pairs and of the chance alignments, we first examine the distributions of angular separation for all the pairs; this is shown in the left panel of \cref{fig:globalreal}.  Although our initial search goes out to an angular separation of \SI{3600}{\arcsec}, the plots only extend out to \SI{400}{\arcsec} to better reveal two main features: a sharp peak at low separations ($\rho < \SI{20}{\arcsec}$) and a steadily increasing distribution of pairs at higher separations.  There is also a sharp drop to zero at very short separations ($\rho < \SIrange[range-phrase={-}]{1}{2}{\arcsec}$, not visible in the left panel of \cref{fig:globalreal}) due to the resolution limits of our initial search catalog, which is effectively that of the Gaia DR2 catalog.  The steadily increasing distribution at large angular separations represents the chance alignment population, while the peak at small angular separations represents the distribution of real pairs.

Further evidence that the peak represents real pairs is seen in the middle and right panels of \cref{fig:globalreal}, which show the distribution of proper motion differences for the two subsets of pairs separated by the blue line in the left panel of \cref{fig:globalreal}.  The middle panel shows the proper motion difference distribution for the pairs found to the left of the blue line ($\rho < \SI{20}{\arcsec}$), which should be mostly true binaries.  The right panel displays the same but for the pairs to the right of the blue line ($\rho > \SI{20}{\arcsec}$), which should be primarily chance alignments.  The proper motion difference distribution for the close pairs shows what we would expect from physical binaries: these pairs have near-identical proper motions and their proper motion differences are thus heavily concentrated near the origin.  The small dispersion about the origin is consistent with the astrometric errors in the Gaia DR2 proper motion measurements.  On the other hand, pairs with large angular separations show what one would expect from random pairings of objects in the plane of the sky, with a very broad distribution of proper motion differences.  These plots, however, do not represent the true distributions of real pairs and chance alignments because each subset contains a mix of both types of CPM pairs.  In order to determine the probability distribution functions for the true binaries and chance alignments, we need to cleanly separate out the true binaries from the chance alignments.

\begin{figure*}
\centering
\subfigure{
\includegraphics[scale=0.37]{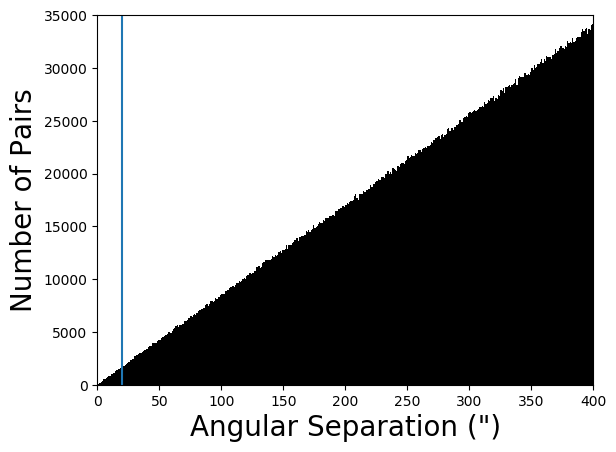}}
\subfigure{
\includegraphics[scale=0.37]{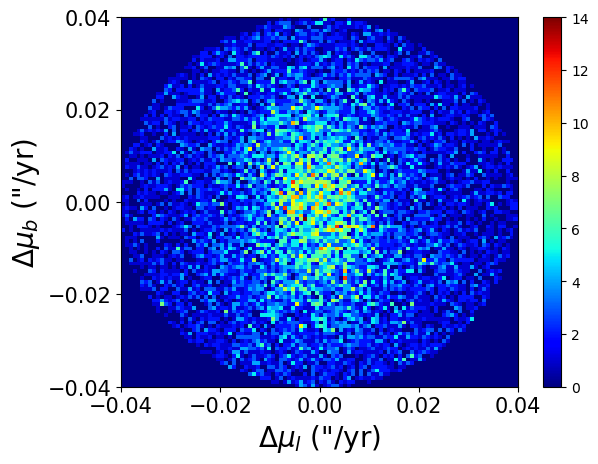}}
\subfigure{
\includegraphics[scale=0.37]{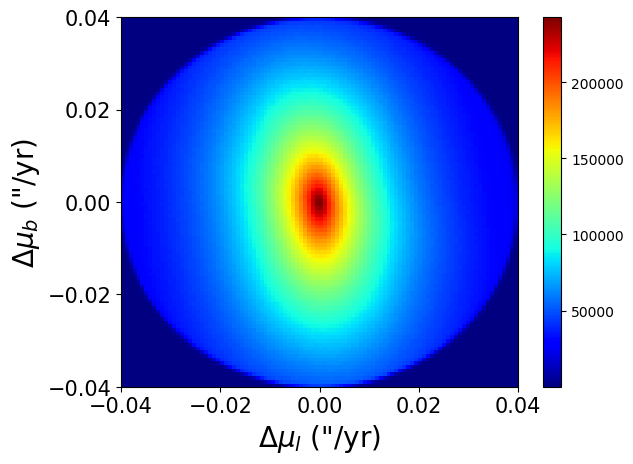}}
\caption{\label{fig:globalrand}As in \cref{fig:globalreal}, but for the randomized cross-match. Notice that unlike in \cref{fig:globalreal}, the pairs at low separations here have a similar distribution of proper motion differences (middle panel) as those at large separations (right panel), confirming that all pairs in this subset are chance alignments.
}
\end{figure*}

To do this, we follow the suggestion of \citet{2007lepine} and create a second sample of possible pairs, but one that would be, by design, completely devoid of physical pairs.  We create this random catalog by copying the original catalog and shifting all of the stars by \SI{4200}{\arcsec} in declination. This randomizes the positions of the stars and gets rid of the real binaries while retaining most of the information about the local sky density and local proper motion distribution of the stars, both of which critically affect the local statistical distribution of angular separations and proper motion differences. We then rerun our search algorithm, but matching the positions and proper motions of stars in the original catalog to those in the random catalog.  The left panel of \cref{fig:globalrand} shows the distribution of angular separations for this random (and true-binary free) cross-match.  Note that the peak at low separations is now absent, which is consistent with an absence of real binaries.  The distribution of proper motion differences is also very revealing: pairs with short angular separations ($\rho < \SI{20}{\arcsec}$; middle panel in \cref{fig:globalrand}) now show no peak near 0, and their distribution of proper motion differences is now very spread out, while pairs with large angular separations ($\rho > \SI{20}{\arcsec}$; right panel in \cref{fig:globalrand}) show a distribution similar to that of the larger separation pairs in \cref{fig:globalreal}. One can see that the distribution of proper motion differences for the short-separation and large separation pairs now appears very similar, which suggests that both subsets represent chance alignments.  In addition, the distribution continues to steadily increase going out to our search limit of $1\degree$.  Thus, using this random catalog, we can independently determine the distribution of the chance alignment population.  We can then scale this distribution to the catalog containing real pairs and subtract off the chance alignments, revealing the statistical distribution of the true (physical) binaries.

\subsection{Searching for Binaries by Subsets}

\begin{figure}[ht]
\centering
\subfigure{
\includegraphics[scale=0.5]{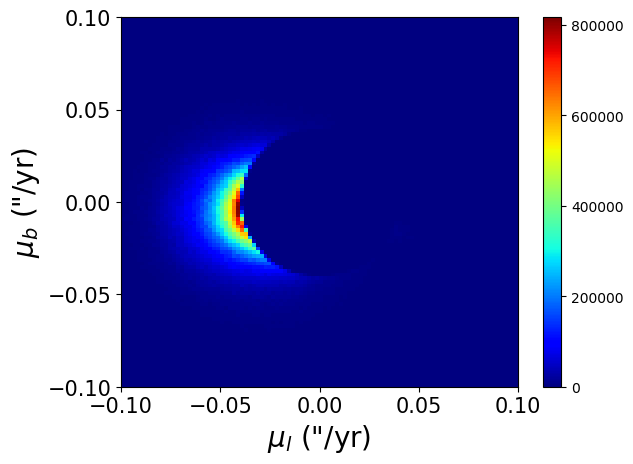}}
\subfigure{
\includegraphics[scale=0.5]{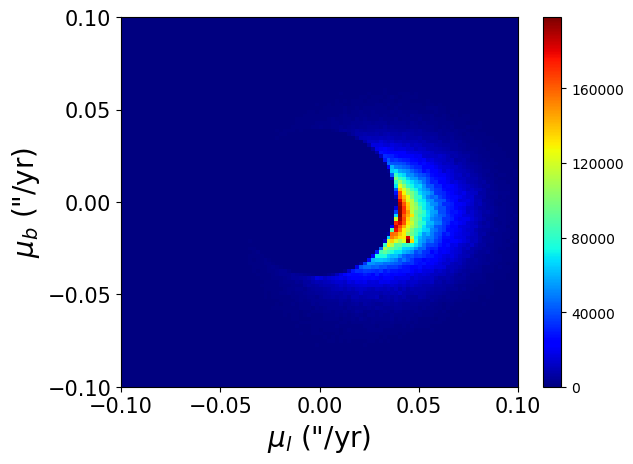}}
\caption{\label{fig:Pmdists12}Distribution of the proper motion values for the common proper motion pairs located towards the Galactic Center (e.g., near $l=0,\, b=0$; top panel) and in the direction of the Sun's motion around the Galaxy ($l=90,\, b=0$; bottom panel).  The asymmetric drift of local stars relative to the Sun's rest frame causes a dramatic asymmetry in the general distribution of proper motions, with strong dependence on location on the sky. This demonstrates why each sky sector needs to be analyzed separately.}
\end{figure}

The search described above can be done globally for the entire data set or for any determined subset representing specific regions of the sky, specific ranges of proper motions, or any other parameter.  As it turns out, the density of possible pairs varies considerably across the sky and is also strongly dependent on the mean proper motion of the pair; this is especially true for the chance alignments.  One major factor for this is the intrinsic distribution of proper motion components $\begin{pmatrix}\mu_l & \mu_b\end{pmatrix}$  for high proper motion stars in general; this can be seen in \cref{fig:Pmdists12} which shows the proper motion distribution for primary stars located in two different regions of the sky: towards the galactic center ($l=0$, $b=0$, top panel) and in the apex of the Sun's motion around the Galaxy ($l=90$, $b=0$, bottom panel).  Three different effects are illustrated in these plots.  

\subsubsection{Splitting by Proper Motion Magnitude}
\begin{figure*}
\centering
\subfigure{
\includegraphics[scale=0.36]{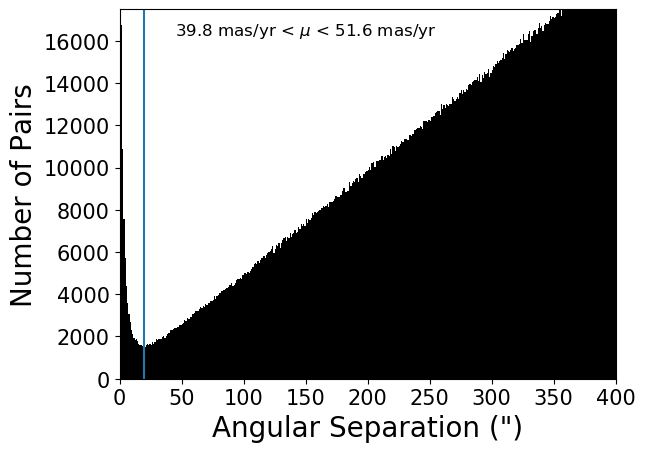}}
\subfigure{
\includegraphics[scale=0.36]{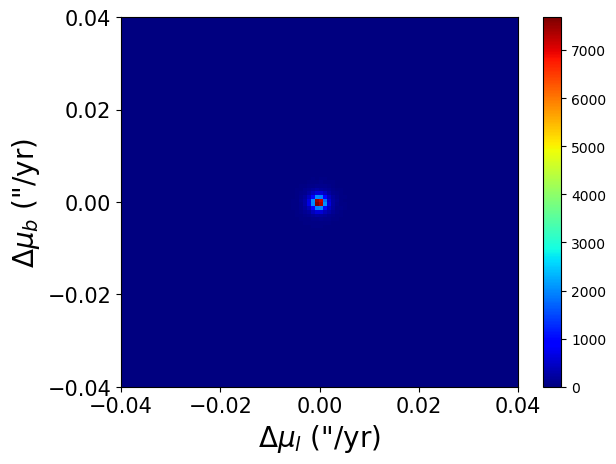}}
\subfigure{
\includegraphics[scale=0.36]{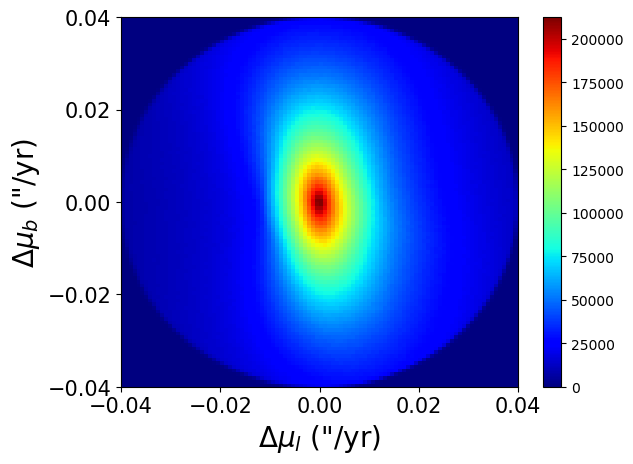}}
\subfigure{
\includegraphics[scale=0.36]{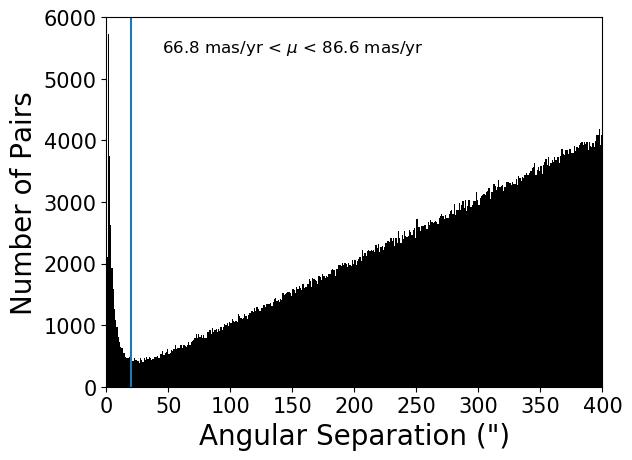}}
\subfigure{
\includegraphics[scale=0.36]{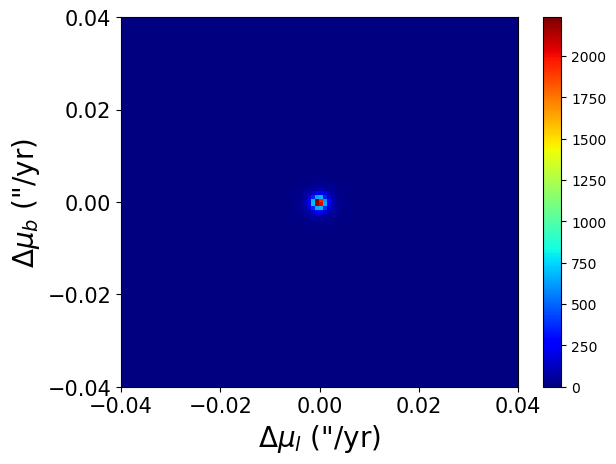}}
\subfigure{
\includegraphics[scale=0.36]{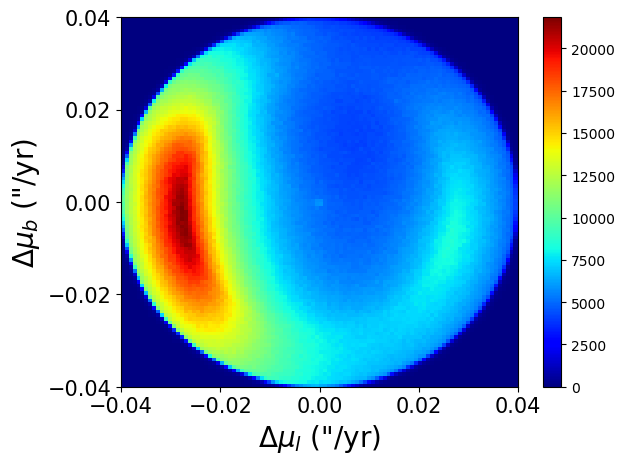}}
\subfigure{
\includegraphics[scale=0.36]{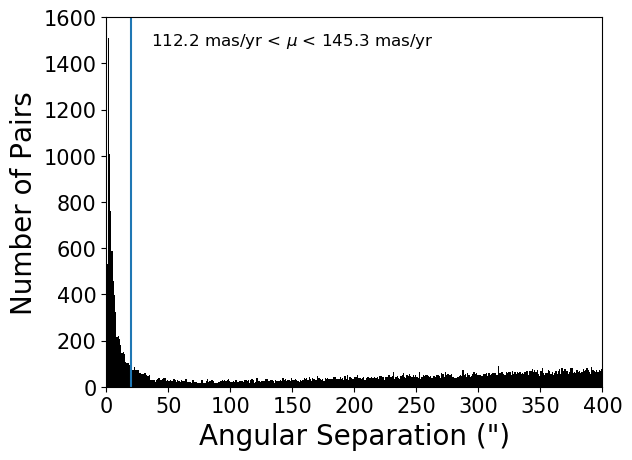}}
\subfigure{
\includegraphics[scale=0.36]{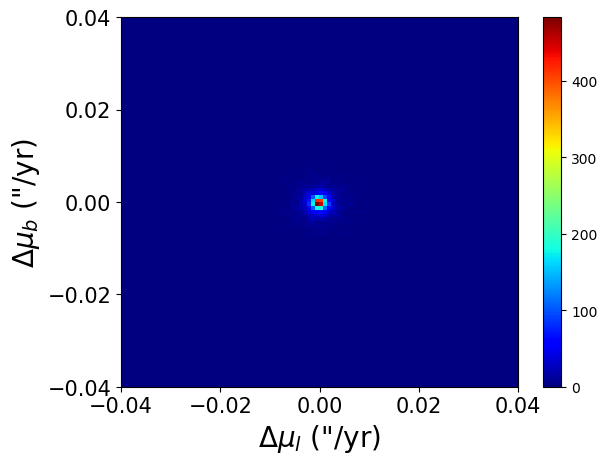}}
\subfigure{
\includegraphics[scale=0.36]{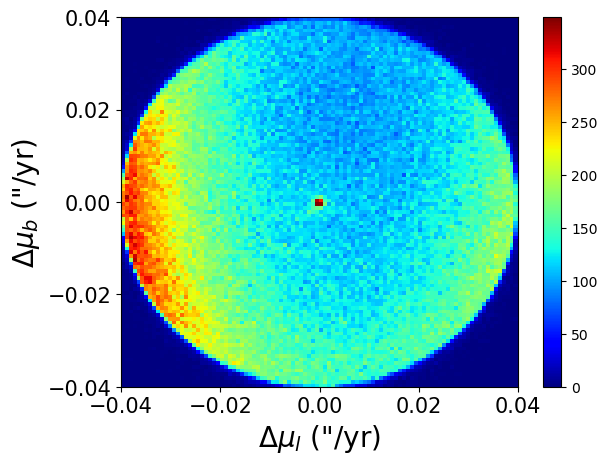}}
\caption{\label{fig:pmsplit}Angular separation and proper motion difference distributions for different proper motion bins.  The top row show the distribution for pairs with total proper motions in the range  39.8<$\mu$<51.6 mas/yr; the central and bottom rows shows the bins for pairs with proper motions 66.8<$\mu$<86.6 mas/yr, and 112.2<$\mu$<145.3 mas/yr respectively.  The amount of chance alignments and real pairs decrease as a function of proper motion.  However, the number of chance alignments drops at a higher rate than the real pairs, allowing for easier identification of real binaries at higher proper motions.}
\end{figure*}

First, the number of stars generally decreases as the magnitude of the proper motion increases.  At lower proper motions, there are more stars and thus more possible chance alignments.  At higher proper motions, the number density of stars (and thus of chance alignments) decreases significantly.  Therefore, we split our sample based on the proper motion magnitude of the brighter star (i.e. the "primary") as determined by their $G$ or $V$ magnitude.    The use of the G or V magnitude depends on whether the primary was a Gaia or only SUPERBLINK source.  We used six bins of proper motion magnitude starting at \SI{39.8}{\mas\per\year} and increasing in steps of \SI{0.1125}{dex}, so that the six bins have edges \SIlist{39.8;51.6;66.8;86.6;112.2;145.3}{\mas\per\year}, with the last bin including all pairs with proper motions \SI{> 145.3}{\mas\per\year}.  \Cref{fig:pmsplit} shows the angular separation and proper motion difference distribution for stars in three of these proper motion bins. As in Figs.1-2, we define two subsamples of pairs with small (<20") and large (>20") angular separations, and plot the distribution of proper motion difference for each subsample (middle and right panels). Again, we see that the latter subset is always dominated by chance alignments.  It also shows that as proper motion increases (from top to bottom in the figure), one observes that the ratio of the number of real pairs to the number of chance alignments increases, allowing for easier identification of true binaries at larger angular separations. 

\subsubsection{Splitting by Sky Vectors}
Examination of \cref{fig:Pmdists12} shows two other potential dependencies: \begin{enumerate*}[label=(\arabic*)] \item the location of the pair on the sky and \item the orientation of the proper motion vector. \end{enumerate*} These effects happen because of the "asymmetric drift" effect, which causes local stellar motions to show a preferred direction in their space motion relative to the Sun.  The two sky regions shown as examples in \cref{fig:Pmdists12} -- one towards the Galactic Center (top panel) and the other towards the apex of the Sun's motion around the Galaxy (bottom panel of \cref{fig:Pmdists12}) -- have proper motion distributions that are flipped from one another.  In addition, the distribution is not uniform in all directions. 

To account for these potential biases, we further split the sample based on the orientation of the proper motion vector (8 octants) and the location on the sky (6 sectors) of the primary star.  To split our sample of pairs based on their location on the sky, we took the coordinates of the primary (brighter) star in each pair and calculated its Galactic Cartesian coordinates $X_g, Y_g, Z_g$.  Using these, we split the sample into 6 different sectors based on which coordinate axis the primary was closest.  These sectors correspond to the axes pointing away and towards the Galactic center (+$\hat{x}$ ,-$\hat{x}$) in the direction of and opposite to the motion of the Sun (+$\hat{y}$ ,-$\hat{y}$) and up and down out of the disk of the Galaxy (+$\hat{z}$ ,-$\hat{z}$).

After all the pairs in the sample are sorted into these six sky sectors, we select a local coordinate system for each sector to represent the proper motions of all the stars in the sectors. For the four sectors along the Galactic equator (+x,-x,+y,-y) we simply use the Galactic proper motion vectors ($\mu_l,\mu_b$). For the two sectors near the Galactic poles (+z,-z) however, we adopt a different system in order to avoid pole effect confusion in the proper motion vector orientations.  In those two sectors, we use a spherical coordinate system that is tilted by 90 degrees to the Galactic system. The two angular coordinates in this reference frame are labeled $r$ and $s$, and are analogs of $l$ and $b$ except they correspond to a coordinate system that has its North pole pointing in the directions of the Galactic Center.   This is done to simplify the analysis of the proper motion differences, as the proper motions ($\mu_r,\mu_s$) of the stars in these new coordinates point in the same direction in each region instead of wrapping around the two pole regions.  

\subsubsection{Splitting by Proper Motion Octants}
\label{sec:muOctantSplit}
\begin{figure}[ht]
\centering
\includegraphics[scale=0.5]{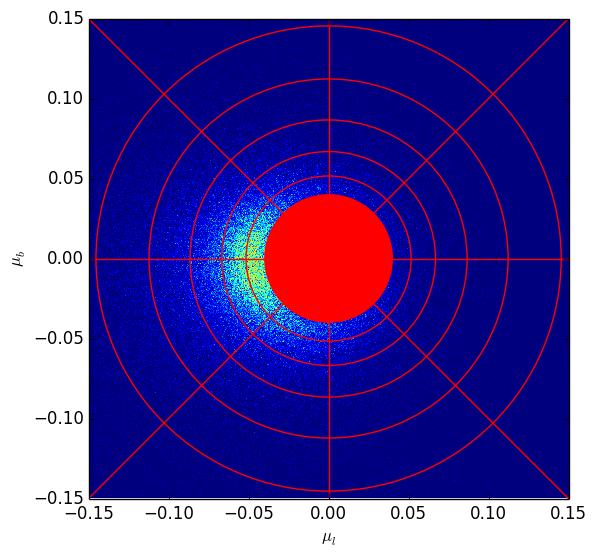}
\caption{\label{fig:showpmoct}Proper motion distribution for stars in the sky sector in the direction of the Galactic Center.  The red circle in the middle represents the area of low proper motion space that is excluded from our search for common proper motion pairs by design.  The red lines and circles show how we separated our six sectors into different regions of proper motion.  The 5 red circles form 6 annuli of proper motion magnitude starting from \SI{39.8}{\mas\per\year}.  The red lines divide proper motion space further into 8 proper motion directions.  Combined, these form 48 different areas to be examined per sector, leading to a total of 288 different areas examined by our code.}
\end{figure}

Finally, the sample is split into eight octants of proper motion direction using the $\mu_l$ and $\mu_b$ proper motion values of the primary star for the first four sky sectors and the $\mu_r$ and $\mu_s$ proper motion values for the two pole sectors.  These octants are in addition to the bins in proper motion magnitude described above.  \Cref{fig:showpmoct} shows the results of these proper motion magnitude and orientation bins for one sky sector.  The red circle in the middle represents the area with proper motion less than \SI{39.8}{\mas\per\year}, which is excluded from the search by design.  Each box represents an area of proper motion space that we characterized and analyzed independently using our Bayesian analysis; this creates 48 different ``boxes'' in proper motion space.  

After all the divisions are performed, we end up with 288 independent bins/sectors/octants on which to perform our analysis.  In order to allow for easier identification, we create a simple coordinate system to identify a specific region in the form (S\#,B\#,O\#) corresponding to the sky sector, proper motion bin and proper motion vector octant.  Each of these areas (location, proper motion magnitude and proper motion vector) are explained in Table \ref{tab:Sectortab}, Table \ref{tab:Propermotionmagtab} and Table \ref{tab:Propermotionorientationtab} respectively.  
\begin{deluxetable}{ll}
\tablewidth{700pt}
\tabletypesize{\scriptsize}
\tablecaption{\label{tab:Sectortab}Description of the Sky sectors used to split the sample.  Coordinates of primary star as determined by magnitude are used for this.}
\tablehead{
\colhead{Sector} & \colhead{Direction on the Sky}
}
\startdata
S1 & Towards the direction of the Galactic Center\\
S2 & Towards the direction of the Sun's motion around the Galaxy\\
S3 & Towards the direction of the Galactic Anti-Center\\
S4 & Towards the direction opposite of the Sun's motion around the Galaxy\\
S5 & Up out of the Galactic Plane\\
S6 & Down out of the Galactic Plane\\
\enddata
\end{deluxetable}

\begin{deluxetable}{ll}
\tablewidth{700pt}
\tabletypesize{\scriptsize}
\tablecaption{\label{tab:Propermotionmagtab}Description of proper motion bins used to split the sample.  Proper motions of the primary stars are used to split the sample.}
\tablehead{
\colhead{Bin} & \colhead{Proper motion range}
}
\startdata
B1 & 39.8<$\mu$<51.6 mas/yr\\
B2 & 51.6<$\mu$<66.8 mas/yr\\
B3 & 66.8<$\mu$<86.6 mas/yr\\
B4 & 86.6<$\mu$<112.2 mas/yr\\
B5 & 112.2<$\mu$<145.3 mas/yr\\
B6 & $\mu$>145.3 mas/yr\\
\enddata
\end{deluxetable}

\begin{deluxetable}{ll}
\tablewidth{700pt}
\tabletypesize{\scriptsize}
\tablecaption{\label{tab:Propermotionorientationtab}Description of how proper motion vectors were used to split the sample.  Proper motions of the primary stars were used to split the sample.}
\tablehead{
\colhead{Octant} & \colhead{Proper motion orientation}
}
\startdata
O1 & |$\mu_l$| > |$\mu_b$| and $\mu_l, \mu_b$ > 0.0\\
O2 & |$\mu_l$| < |$\mu_b$| and $\mu_l, \mu_b$ > 0.0\\
O3 & |$\mu_l$| < |$\mu_b$| and $\mu_l$ < 0.0 and $\mu_b$ > 0.0\\
O4 & |$\mu_l$| > |$\mu_b$| and $\mu_l$ < 0.0 and $\mu_b$ > 0.0\\
O5 & |$\mu_l$| > |$\mu_b$| and $\mu_l$ < 0.0 and $\mu_b$ < 0.0\\
O6 & |$\mu_l$| < |$\mu_b$| and $\mu_l$ < 0.0 and $\mu_b$ < 0.0\\
O7 & |$\mu_l$| < |$\mu_b$| and $\mu_l$ > 0.0 and $\mu_b$ < 0.0\\
O8 & |$\mu_l$| > |$\mu_b$| and $\mu_l$ > 0.0 and $\mu_b$ > 0.0\\
\enddata
\end{deluxetable}

\subsection{Determining the Bayesian Probability Functions}
\subsubsection{Finding $P(\rho|B)$ and $P(\rho|\bar{B})$}

\begin{figure}[ht]
\centering
\includegraphics[width=3.3in]{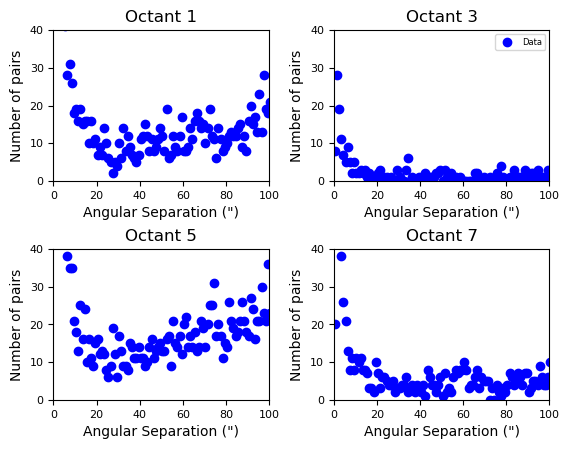}
\caption{\label{fig:Sepdistnomod} Examples of angular separation distributions for pairs in 4 of the 288 independent bins/sectors/octants.  These pairs are from proper motion bin B3 and sky sector S2.  All plots show a peak at low separations representing real pairs and then a steadily increasing trend at higher separations representing chance alignments - note that varying level of chance alignments in the various octants.  }
\end{figure}

Figure \ref{fig:Sepdistnomod} shows 4 examples of the angular separation distribution for pairs in the proper motion bin of 66.8 to 86.6 mas/yr (B3) and in the sector of the sky pointing towards the direction of the Sun's motion (S2).  The four different plots represent four different bins of proper motion directions (octants).  In order to determine $P(\rho|B)$ and $P(\rho|\bar{B})$, we need to infer the statistical distributions of both the chance alignments and the real pairs.  Unfortunately, the two distributions overlap.  However, it is possible to infer both distributions by obtaining an independent estimate of the distribution of chance alignments.

\begin{figure}[ht]
\centering
\includegraphics[width=3.3in]{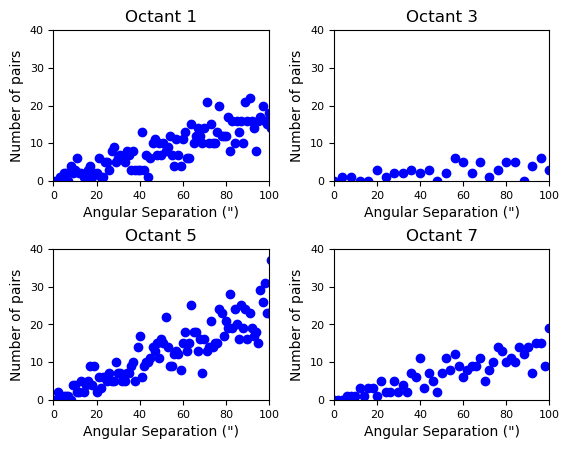}
\caption{\label{fig:Sepdistchance}Examples of angular separation distributions for pairs in 4 of the 288 independent bins/sectors/octants, this time for the randomized catalog.  These pairs are from proper motion bin B3 and sky sector S2.  There is no peak at low separations, which confirms that the randomized set does not contain any real pairs.  Only the steady rise from the chance alignments can be seen.  Number of points is based on the number of pairs in the entire range.}
\end{figure}

Our method to independently map the distribution of chance alignments uses the random catalog we created, as described above in Section 3.1.  We apply the same binning by sector and proper motion to our randomized catalog, and for each bin, we obtain the distribution of angular separations, which now shows only chance alignments.  Figure \ref{fig:Sepdistchance} shows that same areas shown in Figure \ref{fig:Sepdistnomod} from this random catalog.  Notice that the peaks at low separations are now gone, which confirm that the distributions represent only chance alignments.  To model each distribution of chance alignments as a function of angular separation, we fit the cumulative distributions with either a linear or quadratic model, whichever one minimized the chi-squared value.  Once normalized, this model represents $P(\rho|\bar{B})$, which allows us to subtract off the chance alignment trends from the histograms of the true catalog and get the residuals, which represent the distribution of real pairs.

\begin{figure}[ht]
\centering
\includegraphics[width=3.3in]{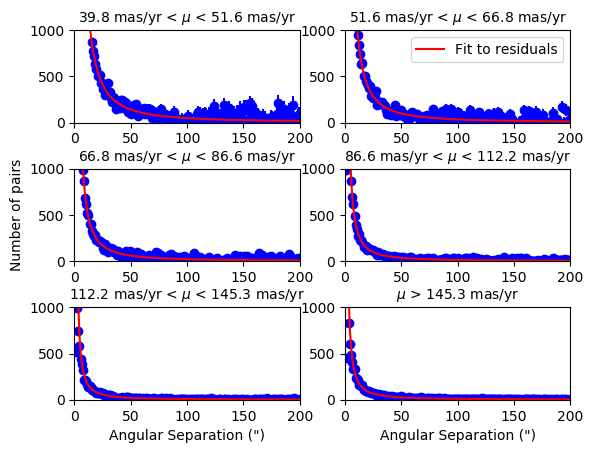}
\caption{\label{fig:Sepdistreal}Combined distribution of angular separations for the real binary models derived for each of the six proper motion magnitude bins.  Note how the mean angular separation becomes shorter at lower proper motions.  Red line represent power law fit of the data.}
\end{figure}

From this method, we derive the statistical distribution of real pairs for each of the 288 sectors/bins/octants.  Our assumption is that the distribution of angular separation for common proper motion binaries should be independent of the sky sector and of the proper motion orientation octant.  On the other hand, the distribution likely varies with proper motion magnitude because proper motion magnitude is correlated with distance for nearby stars.  Therefore, we apply the following procedure:  for each bin of proper motion magnitude, we combine the inferred distribution of real pairs for all 48 sky sectors and proper motion orientation octants.  This generates six independent statistical distributions, which map angular separation for each proper motion magnitude bin.  Figure \ref{fig:Sepdistreal} displays the combined real pair distribution for each proper motion bin, with a power law fit to each one.  This fit, normalized by the integrated value of the function from $2"$ to $3600"$, represents $P(\rho|B)$.  

\begin{figure*}[ht]
\centering
\includegraphics[width=1.0\textwidth]{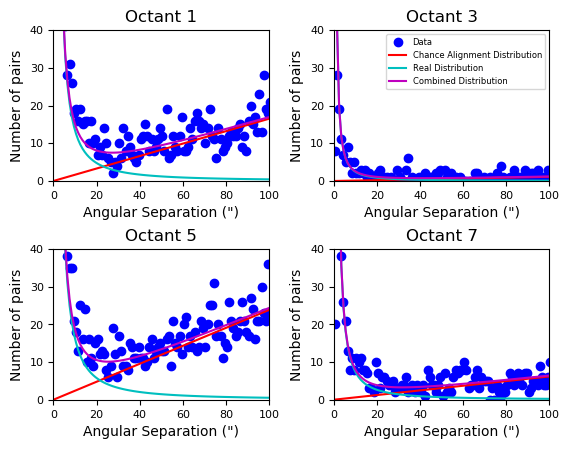}
\caption{\label{fig:Sepdistwithmod}Same plot as Figure \ref{fig:Sepdistnomod}, but now showing the derived models for the angular separation distributions of the real binaries (cyan) and for the chance alignments(red), with the combined distribution shown in purple.}
\end{figure*}

Figure \ref{fig:Sepdistwithmod} shows the same 4 plots of angular separation from the area shown in Figure \ref{fig:Sepdistchance} except now with the real and chance models included on them.  These models are the power law fit which represents $P(\rho|B)$ (cyan line) and the linear or quadratic fit which represents $P(\rho|\bar{B})$ (red line), each scaled to the individual area shown.  The magenta line shows the two models added together to form a combined fit.  


\subsubsection{Finding $P(\Delta\mu_l,\Delta\mu_b|B)$ and $P(\Delta\mu_l,\Delta\mu_b|\bar{B})$}
\begin{figure*}[ht]
\centering
\includegraphics[width=1.0\textwidth]{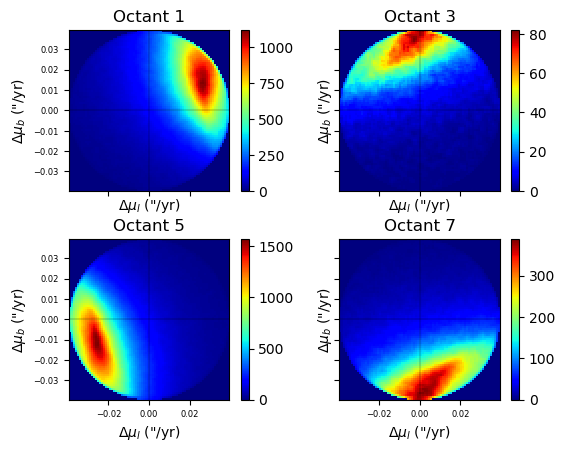}
\caption{\label{fig:Pmdistchance}Distribution of proper motion differences for pairs in the randomized catalog, representing the expected distributions of chance alignments.  The four examples shown here are for the 4 areas represented in Figure 7.  The chance alignment distribution is revealed to be significantly dependent on the orientation of the proper motion vector.  The apparent lines at x=0 and y=0 are to guide the eye and are not real.}
\end{figure*}

To calculate the models for the proper motion differences, we examine 2-D histograms of the distributions of pairs as a function of either $\Delta \mu_l, \Delta \mu_b$ for the four sectors centered along the Galactic equator or $\Delta \mu_r, \Delta \mu_s$ for the two sectors centered on the Galactic poles.  We follow the same procedure as with the angular separations: \begin{enumerate*}[label=(\arabic*)] \item determine the chance alignment distribution from the randomized data, \item subtract it off from the sample containing real pairs and \item determine the real pair distribution from the residuals.  Figure \ref{fig:Pmdistchance} shows the smoothed distribution of $\Delta \mu_l$ and $\Delta \mu_b$ for pairs from the randomized catalog for the same 4 areas in Figure \ref{fig:Sepdistchance}.  These represent the expected distribution of $\Delta \mu_l$ and $\Delta \mu_b$ for chance alignments in each of these areas. \end{enumerate*} As can be seen, the distribution of proper motion differences is not uniform and varies significantly with the orientation of the proper motion vector.  The outer circular edge represents the proper motion difference limit of 40 mas/yr that was imposed in the initial search.  These 2-D histograms are effectively models for the chance alignment distribution $P(\Delta\mu_l,\Delta\mu_b|\bar{B})$.  Due to their level of complexity, we do not attempt to model them with an analytical function, but use the histograms themselves as an empirical model, with the probabilities calculated form the number of pairs in each area divided by the total number of pairs in the histogram.

\begin{figure*}[ht]
\centering
\includegraphics[width=1.0\textwidth]{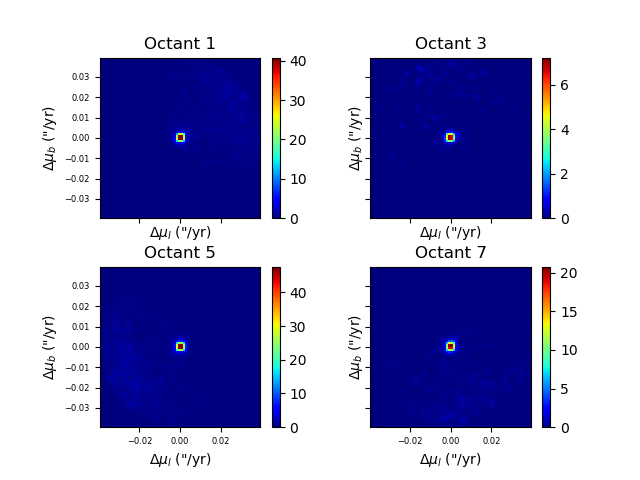}
\caption{\label{fig:Pmdistreal}Proper motion difference distribution for real pairs with separation limits in place.  Same region as Figure 10.}
\end{figure*}

To reveal the distribution of $\Delta \mu_l$ and $\Delta \mu_b$ for the real binaries in the catalog, we set an upper limit on the angular separation and plot the distribution of proper motion differences only for stars within that limit (Figure \ref{fig:Pmdistreal}).  This is what was done previously in Figure \ref{fig:globalreal} to reveal the distribution of real binaries (blue line).  For the two lowest proper motion bins (B1 and B2), the limit was set at $50"$, for bins B3 and B4, $100"$, $250"$ for Bin B5, and then for Bin B6, the limit was set at $400"$.  The limit moves out farther each time because at higher proper motions, the number of chance alignments drops and the number of real pairs rises so we can examine further out with less contamination.  Our previous assumption that the distribution of angular separation for common proper motion binaries should be independent of the sky sector and of the proper motion orientation octant once again works here.  We combine the sky sectors and proper motion orientation octants into six bins where the only difference is proper motion magnitude.  

\begin{figure}[ht]
\centering
\includegraphics[width=3.3in]{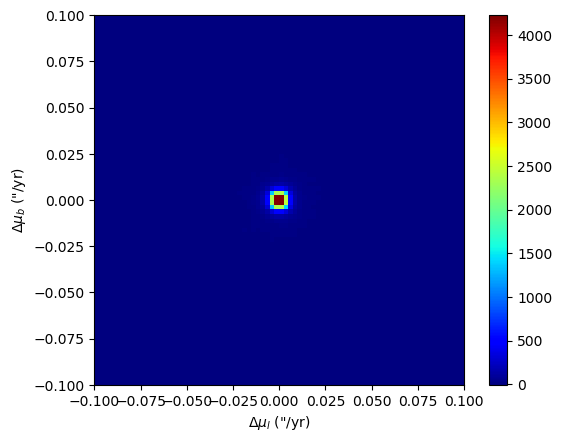}
\caption{\label{fig:Pmsdistmodreal}Model for proper motion difference distribution for real pairs.  This model uses two Gaussian functions.  This plot is for the lowest proper motion bin and is representative of the other five bins.}
\end{figure}

Figure \ref{fig:Pmdistreal} shows the histograms of proper motion differences from then same areas/sectors as in Figure 10, but now for the real binaries (within the imposed angular separation limits).  The peak in the middle around zero clearly represents the real pairs.  There are probably a small number of chance alignments in this sample, however, as evidenced by the random points around the central distribution.  Getting rid of those few remaining chance alignments requires us to create plots from the offset sample also limited to those pairs within the angular separation limits set above.  Most, if not all, of the chance alignment histograms are sparsely populated with many of the bins having zero pairs in them due to the small number of chance alignments with separations less than 100".  To correct this, we smooth those histograms with a box function and do the following: we take all the bins with values less than 1, add them up and divide by the total number of bins.  Any bin with a value less than one is then assigned the value calculated above to ensure calculated probabilities are nonzero.  Taking these chance alignment histograms, we subtract them from the histograms containing the real and chance alignments to get the residuals.  For each proper motion magnitude bin, the residuals from each sector and proper motion direction are added up to form the real pair distribution.  These distributions are then fit with a 2-D model consisting of two Gaussians, both centered at the origin but of different widths.  The best fit is obtained for a narrow Gaussian with dispersion $\sigma = 1.3 mas/yr$ and a broad Gaussian with dispersion $\sigma=4.0 mas/yr$.  The resulting fit and 2-D histograms for the lowest proper motion magnitude bin are shown in Figure \ref{fig:Pmsdistmodreal}.  The rings represent the values of the histogram.  The fits, after being divided by the integral of the function over the entire area, become $P(\Delta\mu_l,\Delta\mu_b|B)$.

\subsubsection{Finding $P(B)$ and $P(\bar{B})$}
The probability priors $P(B)$ and $P(\bar{B})$ are different for each of the 288 subsets but they are derived the same way in each.  For each bin/octant/sector of proper motion magnitude, proper motion orientation and location on the sky, the number of wide binaries ($N_r$) and the number of chance alignments ($N_{ca}$) are derived.  $N_{ca}$ is found by examining the region between 1800" and 3600" in angular separation.  The assumption is that the number of wide binaries in this range is negligible compared to the number of chance alignments.  This allows us to set the number of pairs in that region to be equal to the number of chance alignments multiplied by the integral of the normalized chance alignment model,
\begin{equation}
N_{1800-3600} = N_{ca} * \int_{1800}^{3600} P(\rho|\bar{B}) d\rho
\end{equation}

Solving this equation gave us $N_{ca}$.  The value of $N_r$, on the other hand, if found by fitting the distribution of angular separations with a combined function of the normalized distribution of real pairs times a scalar, $N_r$, and the normalized distribution of chance alignments times $N_{ca}$.  This is represented as the magenta lines in Figure \ref{fig:Sepdistwithmod}.  $P(B)$ and $P(\bar{B})$ can then be rewritten in terms of the the ratio of $N_r$ and $N_{ca}$,

\begin{equation}
P(B) = \frac{N_r}{N_r + N_{ca}} = \frac{1}{1+\frac{N_{ca}}{N_r}}
\end{equation}
\begin{equation}
P(\bar{B}) = \frac{N_{ca}}{N_r + N_{ca}} = \frac{1}{1+\frac{N_r}{N_{ca}}}
\end{equation}

\begin{figure}[ht]
\centering
\includegraphics[width=3.3in]{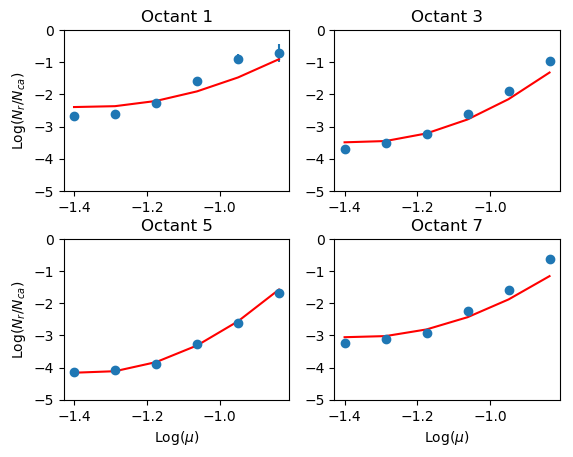}
\caption{\label{fig:NpNc}Ratio of $\frac{N_r}{N_{ca}}$ for the four quadrants examined above as a function of proper motion magnitude and in the direction of Galactic center.  The red line is the fit to the data.}
\end{figure}

It was discovered that the ratio is a function of proper motion magnitude.  Figure \ref{fig:NpNc} shows examples of this ratio for four octants of proper motion orientation, in the direction of the Galactic center and plotted as a function of the proper motion of the primary star in the pair.  The red line represents a quadratic fit to the ratios as a function of proper motion.  This allows us to get a ratio of $N_r$ and $N_{ca}$ and then calculate the priors, P(B) and P($\bar{B}$), for any proper motion of the primary star.

\section{Wide System Identification Method: Second Pass Using Gaia DR2 Parallaxes}
\subsection{Selection of the Second Pass Subset}
After running the 557 million possible pairs through the code, we found 176,896 pairs that have Bayesian probabilities greater than 10\% of being real binaries.  Up to this point, our analysis has consisted of using only the angular separations and proper motion differences of the pairs, which we now refer to as the ``first'' pass.  However, parallax data, if available, can further constrain the probability estimate for the two stars to be physical  binaries.  In this section, we perform a "second pass" by searching for and incorporating the parallax information to the Bayesian analysis.  We run this second pass on the subset of stars with probabilities larger than $10\%$ as determined from the first pass (Section 3 above).  We restrict our parallax analysis only to stars with high probability from the first pass, in order to reduce the computational time.  Tests using the complete catalog show that the vast majority of the pairs ultimately identified as true binaries have first pass probabilities larger than $10\%$. 

The parallax test is based on the premise that if a pair is a true binary, they should be at the same distance.  If their distances are widely different from each other, then they must be chance alignments.  Starting with the 176,896 pairs with first pass probabilities greater than $10\%$, we searched the Gaia DR2 catalog for the parallaxes of both members of each pair.  Reliable parallaxes (parallax errors smaller than $10\%$ of the parallax) could not be found for $\sim33\%$ of the pairs.  In some cases, the parallaxes were listed as negative.  In many pairs, one of the components has a reliable parallax, while the other does not.  In the end, we assembled a subset of 119,390 pairs where both stars have reliable parallaxes; the analysis described below is applied to that subset.

\subsection{Distance Difference Analysis}
\label{sec:deltaDist}
Taking the surviving pairs, we conduct a separate Bayesian analysis, this time based on the difference in distance between the two pairs $\Delta D = D_{pri} - D_{sec}$, where the distances are simply calculated from the parallaxes, $D_{prim} = (\pi_{pri})^{-1}, D_{sec} = (\pi_{sec})^{-1}$. To integrate with the results from the proper motion analysis (i.e. the "first pass"), we consider the probability to be a function not just of the distance difference $\Delta D$ but also of the Bayesian probability calculated in the first pass. If we define $P_{\rho,\mu}$ to be the first pass Bayesian probability such that:
\begin{equation}
P_{\rho,\mu} = P(B | \Delta \mu_l, \Delta \mu_b, \rho),
\end{equation}

then the formula associated with Bayes theorem for our second pass is:

\begin{equation}
P(B|\Delta D,P_{\rho,\mu})=\frac{(P(\Delta D,P_{\rho,\mu}|B)\cdot P(B))}{P(\Delta D,P_{\rho,\mu})}
\end{equation}

where

\begin{equation}
P(\Delta D,P_{\rho,\mu}) = P(\Delta D,P_{\rho,\mu}|B)\cdot P(B) + 
P(\Delta D,P_{\rho,\mu}|\bar{B})\cdot P(\bar{B})
\end{equation}

As in the previous analysis, we need to find the two probability distributions, $P(\Delta D,P_{\rho,\mu}|B)$ and $P(\Delta D,P_{\rho,\mu}|\bar{B})$, and the two priors, P(B) and $P(\bar{B})$, for each individual pair to calculate the final probability that a pair is a real binary.

\subsection{Calculation of the Probability Distributions: $P(\Delta D,P_{\rho,\mu}|B)$ and $P(\Delta D,P_{\rho,\mu}|\bar{B})$}

\begin{figure}[ht]
\centering
\includegraphics[width=3.3in]{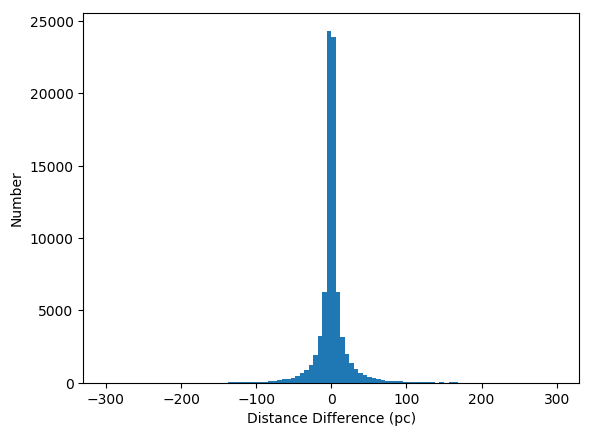}
\caption{\label{fig:realdistdifftot}Distance difference for pairs matched in Gaia dr2 with first run probabilities $>99\%$.  The large peak at 0 represents real pairs while parts of the tails of the distribution represent the chance alignments.}
\end{figure}

The probability distribution $P(\Delta D,P_{\rho,\mu}|B)$ represents the distribution of the distance differences $\Delta D$ for pairs that are real binaries. We extract this probability distribution function by examining the distribution of $\Delta D$ values for the subset of pairs that were identified in the first pass to have probabilities greater than $10\%$ and representing pairs that are most likely to be actual binaries. The distribution of $\Delta D$ values is shown in Figure \ref{fig:realdistdifftot} for those pairs that had $P_{\rho,\mu} > 99\%$. The distribution shows a large peak at 0, which confirms that the majority of the pairs in this first run probability range are indeed physical binaries. Figure \ref{fig:realdistdifftot} however also shows extended wings that suggest a number of pairs in this subset are either not physical binaries, or have significant errors in their Gaia DR2 parallaxes. In order to estimate the true distribution of $\Delta D$ for physical binaries we need to disregard the extended wings, only focusing on the central peak.  Before we can do this, we must first understand the distribution from chance alignments.

The probability distribution $P(\Delta D,P_{\rho,\mu}|\bar{B})$ represents the distribution of $\Delta D$ values for pair that are chance alignments. We extract this probability distribution from our subset by examining the distribution of the distance difference $\Delta D$, for a subset of 61,120 pairs that in the first pass were found to have probability values between $1\%$ and $10\%$, are are thus dominated by chance alignments. The distribution of $\Delta D$ values for this subset is shown in Figure 15. There are two components that can be identified in this Figure.  The first component is the peak at 0, which shows that there are in fact some real pairs in this sample, even with the low first-pass probability range.  The second component is an underlying, broad distribution of chance alignments.  We split the pairs into groups based on their first run probabilities, 10 to 20, 20 to 40, 40 to 60, 60 to 80, 80 to 99 and 99 and up.  We chose not to go below 10$\%$ when deriving the probabilities as the number of chance alignments begin to dominate the distribution and no pairs from that probability range will reach a high probability. 

\begin{figure}[ht]
\centering
\includegraphics[width=3.3in]{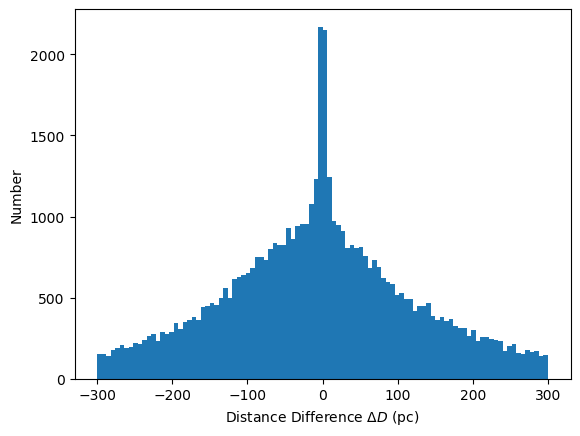}
\caption{\label{fig:10to20distdifftot}Distance difference for pairs matched in Gaia dr2 with first run probabilities between $1\%$ and $10\%$.   The peak at 0 represents real pairs while the wings of the distributions represent chance alignments.}
\end{figure}

\begin{figure}[ht]
\centering
\includegraphics[width=3.3in]{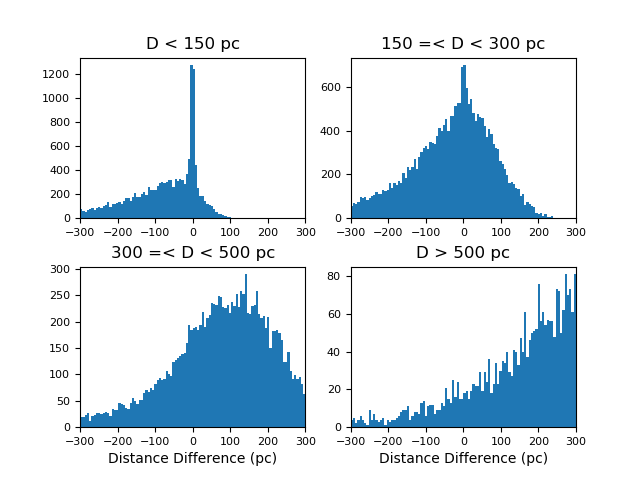}
\caption{\label{fig:10to20distdiff}Distance difference for pairs matched in Gaia dr2 with first run probabilities between $1\%$ and $10\%$, shown for four bins of primary star distance from the Sun (D).  The distribution of chance alignments shifts from left to right depending on distance to the primary star.  In addition, the number of real pairs also decreases with increasing distance.}
\end{figure}

The distribution of $\Delta D$ values for the chance alignments is found be a strong function of primary distance.  Figure \ref{fig:10to20distdiff} shows the distance difference distributions for pairs in the 1\% to 10\% first run probability bin, where the paris are separated into four different distance bins: $D < 150 pc, 150 pc < D < 300 pc, 300 pc < D < 500 pc, D > 500 pc$.  As the primary star's distance increases, the chance alignment distribution shifts from the left to the right and the number of possible physical binaries (the central peak) decreases.  This is because of our high proper motion limit and at higher distances, there is a higher chance that a matched secondary is closer to the Sun rather than farther away.

\begin{figure*}
\centering
\subfigure{
\includegraphics[scale=0.28]{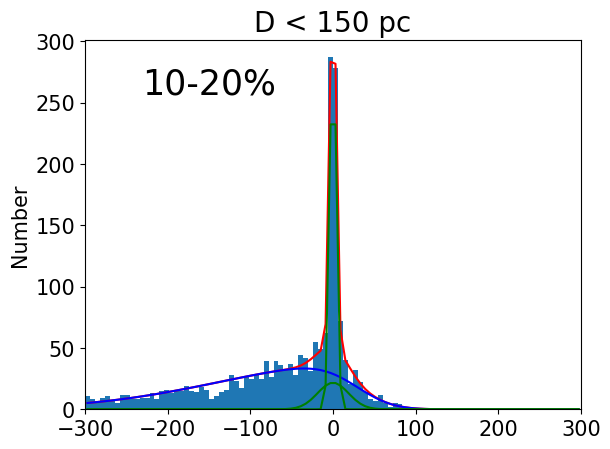}}
\subfigure{
\includegraphics[scale=0.28]{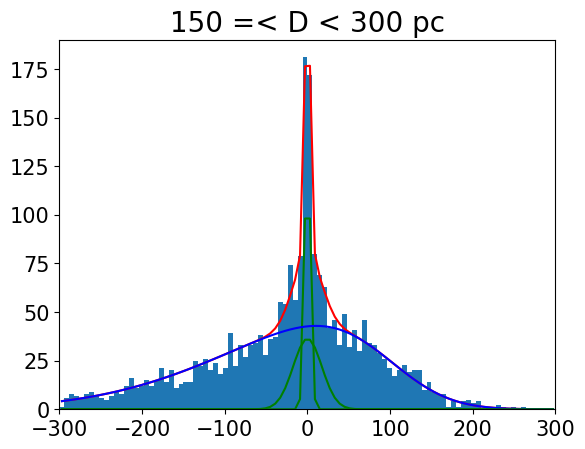}}
\subfigure{
\includegraphics[scale=0.28]{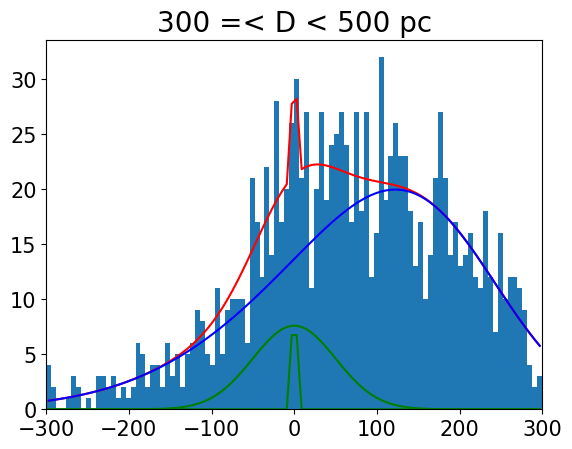}}
\subfigure{
\includegraphics[scale=0.28]{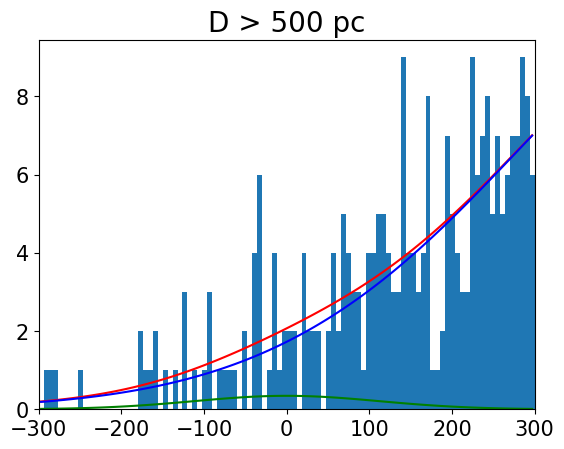}}
\subfigure{
\includegraphics[scale=0.28]{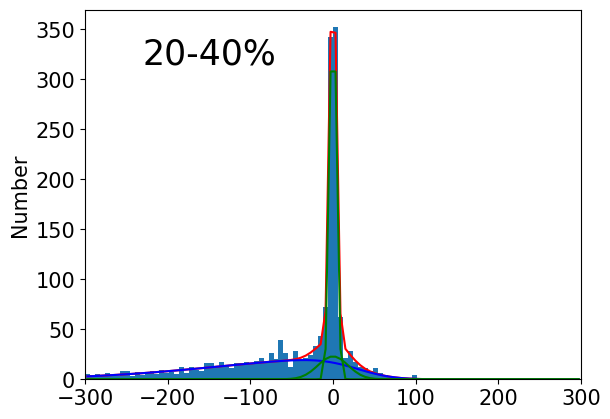}}
\subfigure{
\includegraphics[scale=0.28]{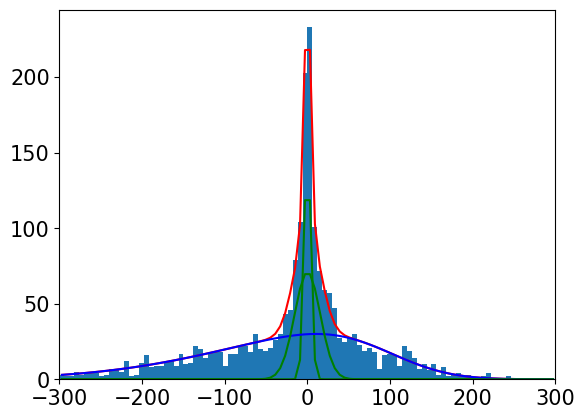}}
\subfigure{
\includegraphics[scale=0.28]{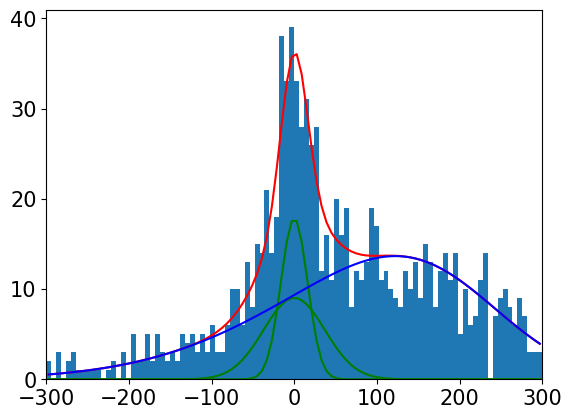}}
\subfigure{
\includegraphics[scale=0.28]{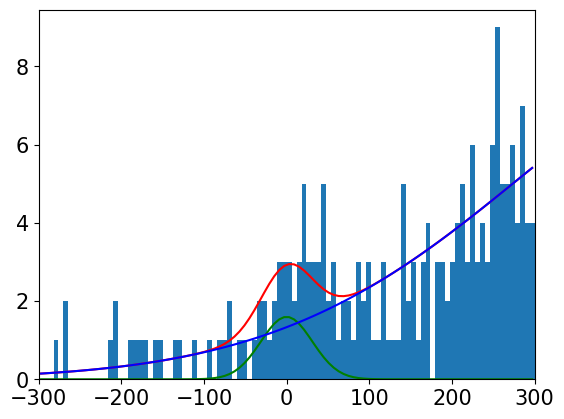}}
\subfigure{
\includegraphics[scale=0.28]{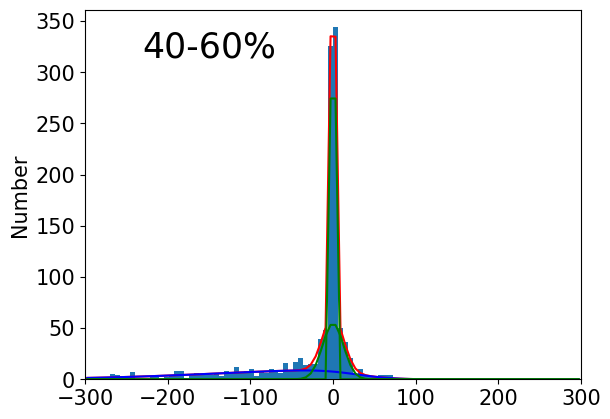}}
\subfigure{
\includegraphics[scale=0.28]{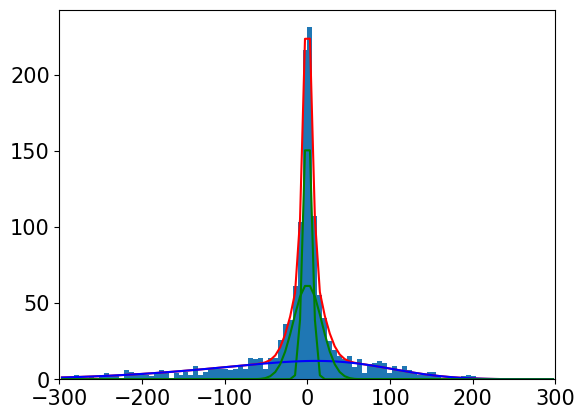}}
\subfigure{
\includegraphics[scale=0.28]{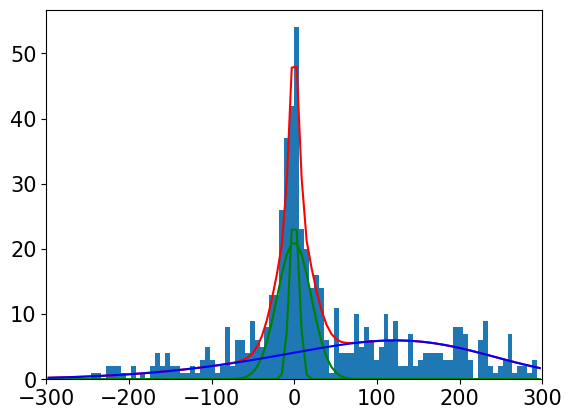}}
\subfigure{
\includegraphics[scale=0.28]{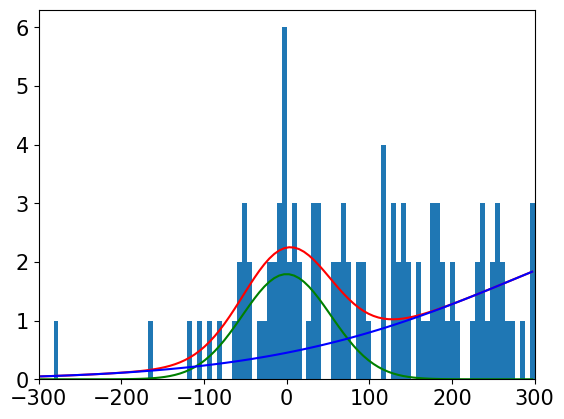}}
\subfigure{
\includegraphics[scale=0.28]{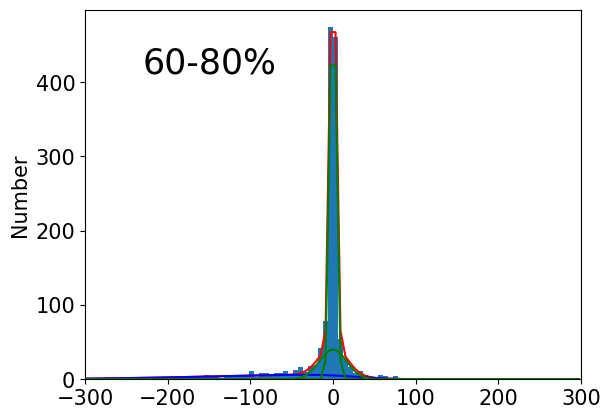}}
\subfigure{
\includegraphics[scale=0.28]{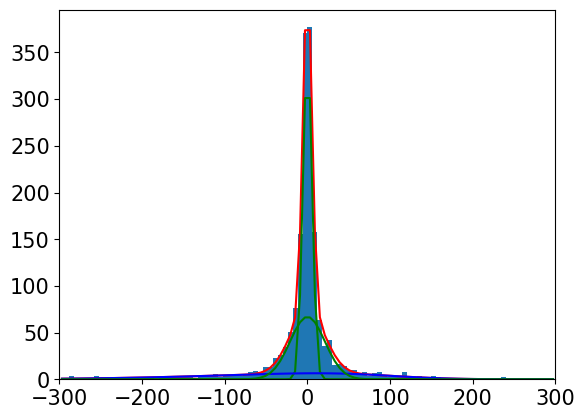}}
\subfigure{
\includegraphics[scale=0.28]{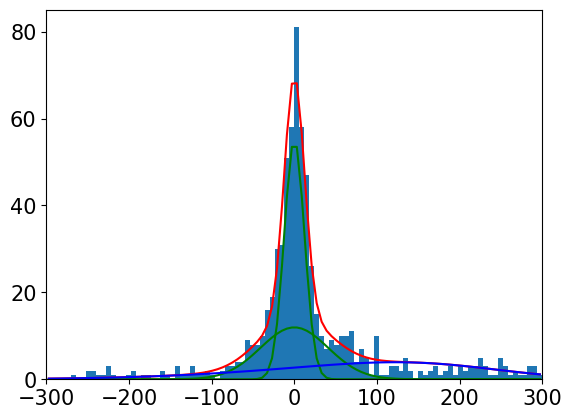}}
\subfigure{
\includegraphics[scale=0.28]{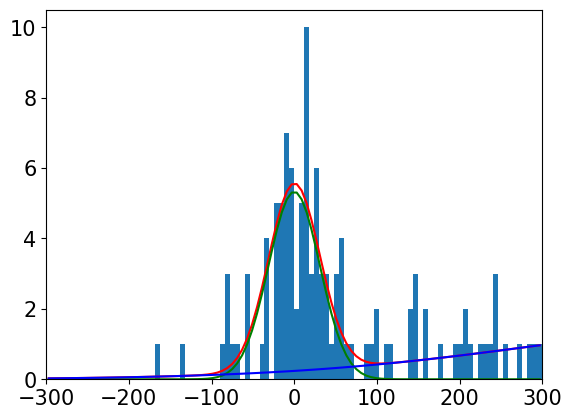}}
\subfigure{
\includegraphics[scale=0.28]{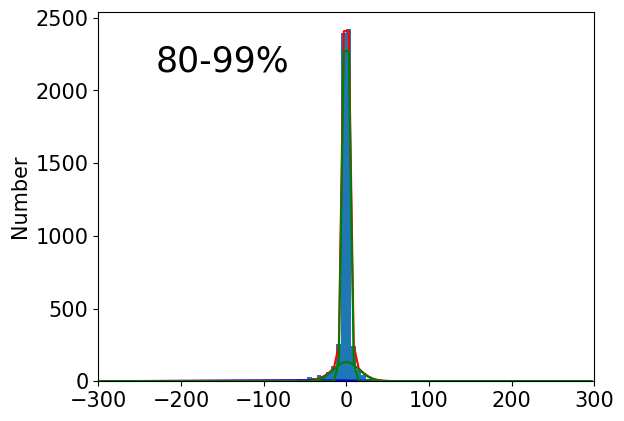}}
\subfigure{
\includegraphics[scale=0.28]{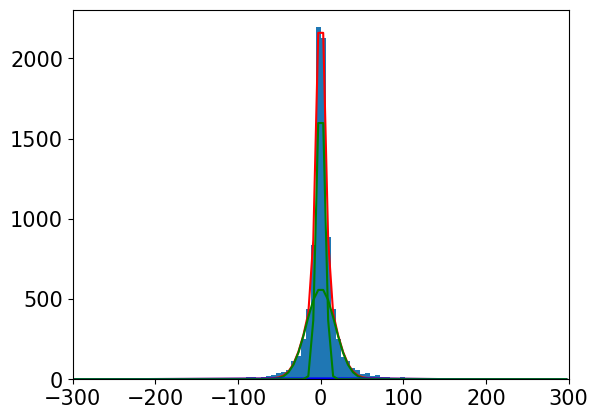}}
\subfigure{
\includegraphics[scale=0.28]{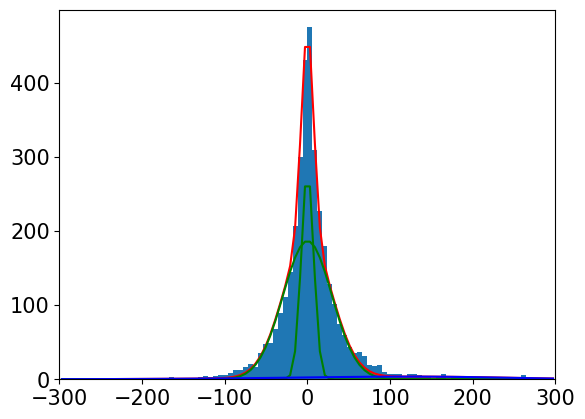}}
\subfigure{
\includegraphics[scale=0.28]{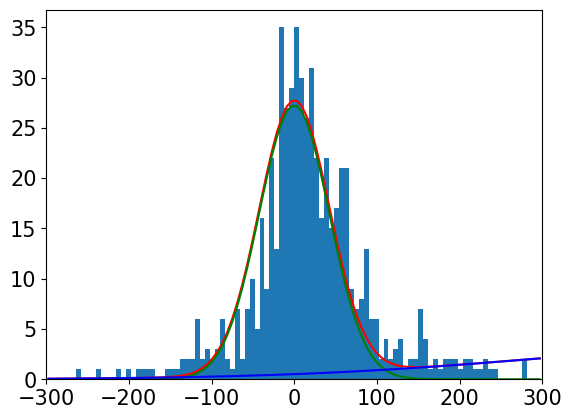}}
\subfigure{
\includegraphics[scale=0.28]{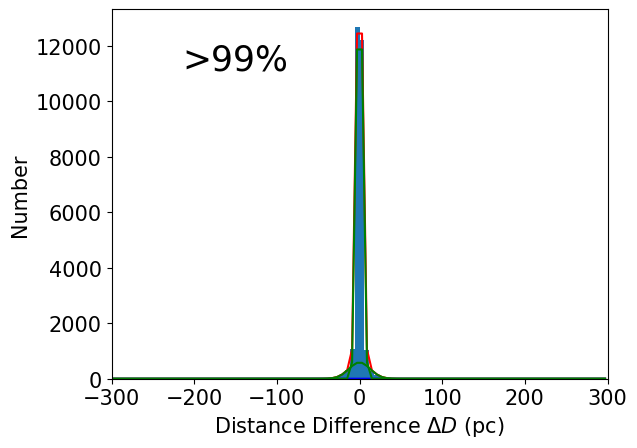}}
\subfigure{
\includegraphics[scale=0.28]{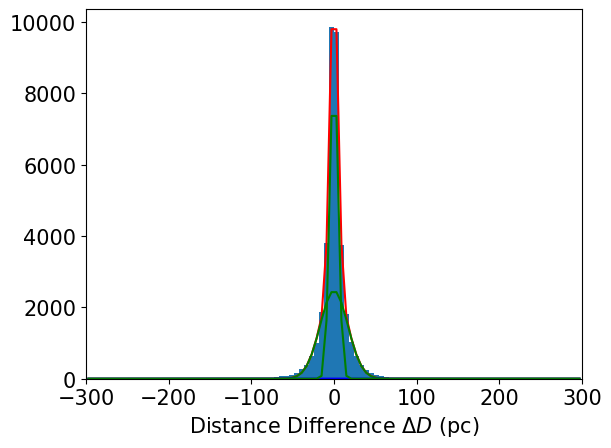}}
\subfigure{
\includegraphics[scale=0.28]{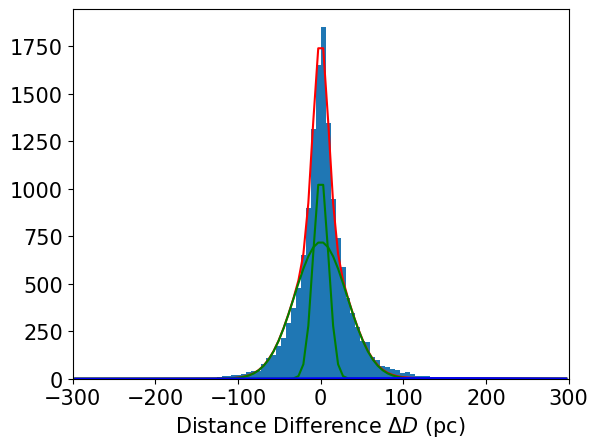}}
\subfigure{
\includegraphics[scale=0.28]{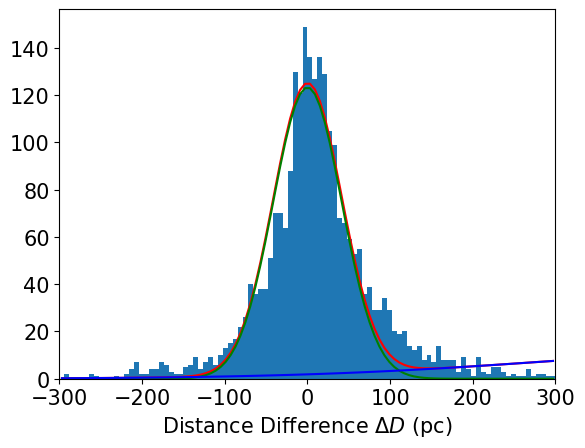}}
\caption{\label{fig:distdiffallprobs}Distance Difference histograms for all first run Bayesian probability bins. Number in the upper left of the left most plot in each row represents the first run probability range being examined in each row.  Lines represent model fits to the histograms.  Blue line represents the chance alignment distribution, green lines represent the real pair distribution, and the red line represents the combined chance and real distributions}
\end{figure*}

To extract a model $\Delta D$ distribution for the chance alignments, we perform a fit of an analytic function to each of the four distance bins.  In each case, the fit excludes pairs with distance differences from -100 to 100 pc to avoid conatmination from physical binaries (the central peak).  After attempting several analytical functions, we find that a skewed Gaussian fit provided reasonably good models for the chance alignments as shown by the blue lines in Figure \ref{fig:distdiffallprobs}.  Once normalized, this function is applied to each probability bin in the same manner as the chance alignment distribution in the first pass where the functional form was multiplied by a scaling factor.  This yields $P(\Delta D,P_{\rho,\mu}|\bar{B})$.

The functional form of the real pairs is still needed however.  To find this, we examine the four distance bins with first run probabilities $>99\%$ and subtract off the chance alignment distribution, leaving behind only the real pairs.  We determined that for the three lower distance bins, a two Gaussian solution matches the real distribution best.  For the highest distance bin (D$>500$ pc), a single Gaussian is used.  We then refit each area with either the single or two Gaussian solution for the real pairs.  This fit is $P(\Delta D,P_{\rho,\mu}|B)$.  Figure \ref{fig:distdiffallprobs} shows the chance alignment model (blue line), real pair model (green line/s) and combined model (red line) for all probability and distance areas in the second pass.  From this figure, one can see that most of the pairs that will have high second pass probabilities come from areas of low distances or high first pass probability.  At low first pass probability, the chance alignments dominate at high distances and are still present at low distances, highlighting the need to characterize them.  

\subsection{Calculation of the Priors for Different Distance Ranges}
\begin{figure}[ht]
\centering
\includegraphics[width=3.3in]{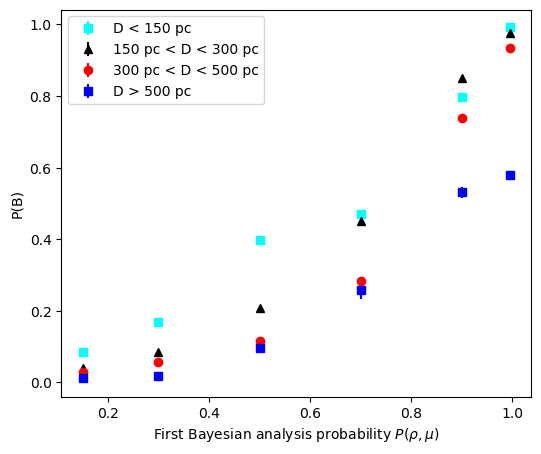}
\caption{\label{fig:pbvsprob}Estimated prior probability of being a binary for the second run, which includes an analysis of the parallax difference between components as a function of the probability of being a binary from the first run.  The priors also depend on the pair's distance; the four distance bins are shown with different symbols.}
\end{figure}

The priors P(B) and $P(\bar{B})$ represent the probabilities for any pair in a particular subset to be either a real binary or not.  To find these priors, we follow a procedure similar to what was used in Section 3.  The priors are calculated separately for each of the 24 different bins, comprising the four different distance bins and the 6 different first run probability ranges.  Taking the number of real and chance alignments from the integrated distributions found above, we derived the priors for each bin and plotted them as a function of first run probability for each of these 24 bins.  Results are shown in Figure \ref{fig:pbvsprob}, which plots the estimated priors as a function of first-run probability, with different symbols denoting the 4 distance bins.  As can be seen, the prior probability depends on the probability from the first run and also significantly depends on the distance of the primary.  The closer a primary star is to us, the higher its prior.  For each probability range, 10\% to 20\%, 20\% to 40\%, 40\% to 60\$, 80\% to 99\% and 99\% and up, we adopt a single value for the prior for each distance range bin, instead of attempting to derive a relationship with distance, which would significantly complicate the problem.  We believe this simplification does not bias the results significantly, as the change in prior over a given first run probability range is an average of 0.15.


\subsection{Unvetted Pairs}
As explained in Section 4, we applied a parallax error cut to the group of pairs that had probabilities of $10\%$ or more after the first pass of the Bayesian analysis that only used angular separation and proper motion difference.  Orginally, 176,896 pairs were in that $10\%$ and up group.  In the second pass of the Bayesian analysis, we only considered pairs in which both components have parallaxes from Gaia or other sources, and where the parallax errors less than 10\% of the parallax itself. This left us with a group of 57,506 pairs that failed this cut.  The pairs that are in this group either have one or more components whose quoted Gaia parallax error is larger than $10\%$ or have one or more component that does not have a parallax value listed in the Gaia catalog and the parallax recovered from the literature is not accurate enough, or one of the components of the pairs has no parallax from Gaia and no parallax from any other source.   We call the pairs the "unvetted" subset, because they are identified based on proper motion and angular separation but are not vetted with parallax data.

\begin{figure}[ht]
\centering
\includegraphics[width=3in]{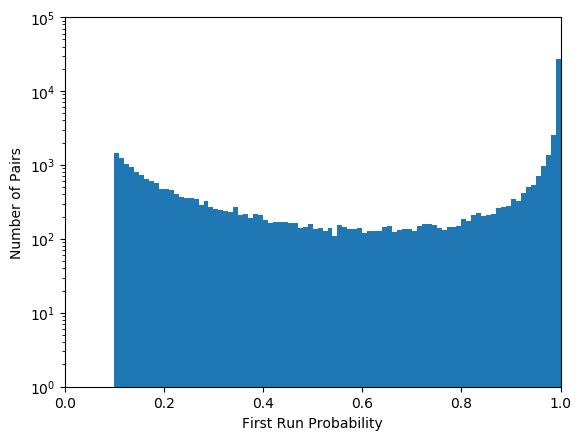}
\caption{\label{fig:failedprobs}First pass Bayesian analysis probabilities for pairs that did not pass the parallax error cut.  Most have high probabilities.  The cut at $10\%$ is due our selection of pairs with probabilities from the first Bayesian analysis greater than that amount}
\end{figure}

\begin{figure*}[ht]
\centering
\includegraphics[width=6in]{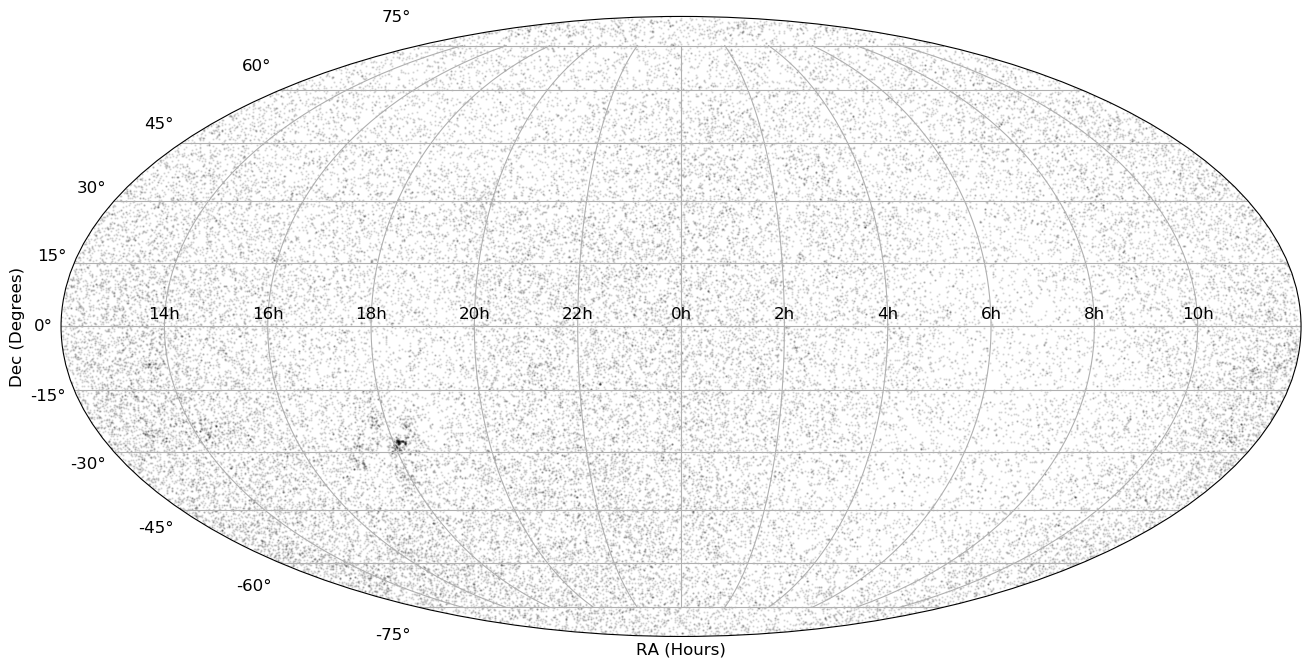}
\caption{\label{fig:failedalsky}All sky distribution of the unvetted pairs, i.e. pairs selected from the first pass, but that did not satisfy the parallax error cut for the second Bayesian analysis.}
\end{figure*}

Figure \ref{fig:failedprobs} shows the first pass probability distribution of these unvetted pairs and shows that the majority of the pairs had very high probabilities of being physically bound systems in this first pass.  We provide the all-sky plot of these pairs in Figure \ref{fig:failedalsky}.  This, combined with the probabilities in Figure \ref{fig:failedprobs}, suggests that most of the pairs in the group are most likely genuine pairs if they have a high first run probability.  One exception is the clump of stars in the direction of the Galactic Center.  There is a well known issue with with the proper motion values of many stars in that area, that are erroneously listed in the Gaia catalog with having large proper motions, and are thus an artifact of the Gaia catalog.  As seen in Figure \ref{fig:failedprobs}, we are confident that most of the pairs in this unvetted subset are real pairs, however, we do caution the user that there are still chance alignments in this subset.  Further vetting of this sample is planned for a future paper.

\section{Results}
\subsection{Catalog of High Probability Wide Binaries From Gaia DR2}
\subsubsection{Catalog of Parallax-Vetted Pairs}
\begin{figure*}[ht]
\centering
\includegraphics[width=6in]{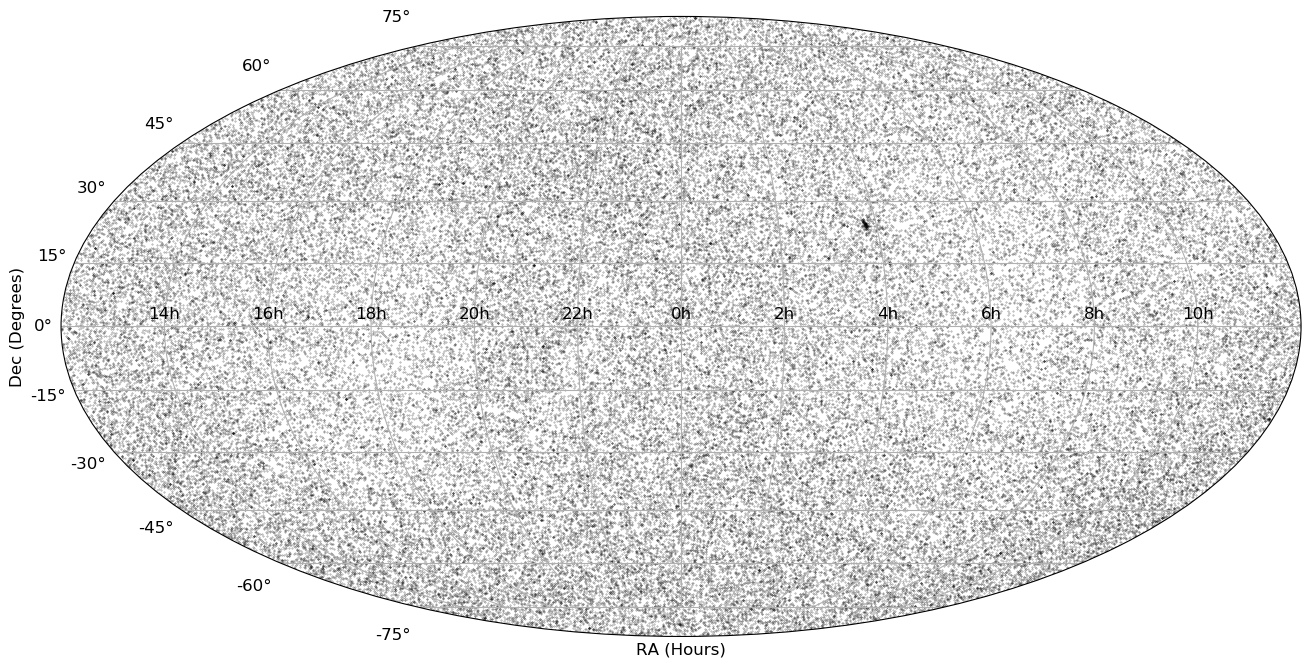}
\caption{\label{fig:vettedallsky}All sky distribution of vetted pairs, i.e. pairs with good Gaia parallaxes and with Bayesian probabilities $>95\%$ after the second pass.}
\end{figure*}

\begin{figure}[ht]
\centering
\subfigure{
\includegraphics[scale=0.5]{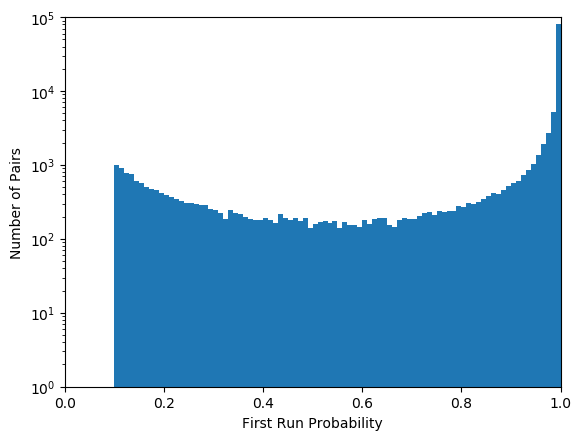}}
\subfigure{
\includegraphics[scale=0.5]{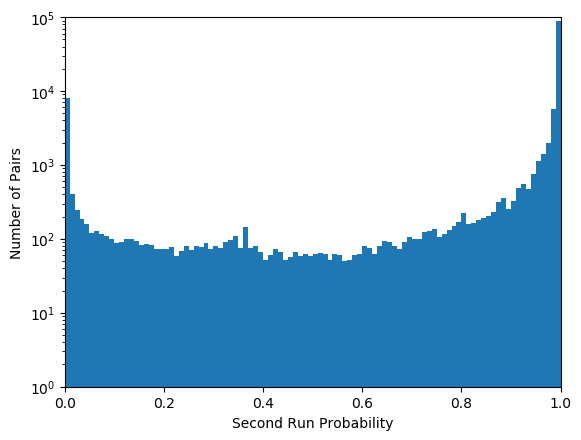}}
\caption{\label{fig:vettedprobs}Distribution of first (top panel) and second pass (bottom panel) probabilities for pairs that have probabilities $>10\%$ that have also passed the parallax error cut.}
\end{figure}

After applying the two Bayesian probability searches described in section 3 (first pass) and section 4 (second pass), we identify 99,203 pairs with probabilities greater than 95\% to be wide binaries.  Of these, we estimate the number of false positives to be about 364.  We calculate this value by summing the individual Bayesian probabilities ($Q_i = 1 - P_i$) that each pair in this high probability subset has of being a chance alignment, where P is the probability of the pair to be a gravitationally bound system.  We show the positions of these wide binaries in Figure \ref{fig:vettedallsky}.  The full catalog which includes these high probability pairs, and also pairs with lower Bayesian probabilities identified in the first and second passes, is presented in three tables.  Figure \ref{fig:vettedprobs} show the probability distributions from the first and second passes of our analysis for all pairs that had probabilities $>10\%$ from the first run and passed our parallax error cut.  Table \ref{tab:primarytab} shows data for all primaries of the matched pairs in GAIA with first pass probabilities $>10\%$ and which also passed our parallax error test and went through the second pass.  This table lists 119,390 pairs of stars.  The table provides the catalog name, Gaia DR2 id, location in RA and DEC in degrees, proper motions in the RA and DEC directions in $mas/yr$, the parallax in mas, the G magnitude, $G_{BP}-G_{RP}$ color, the Gaia radial velocity, if available.  Table \ref{tab:secondarytab} compiles the same information for the secondary stars.  Table \ref{tab:binarytab} gives information about the configuration of the binaries: their angular separation, projected physical separation, G magnitude difference, radial velocity difference if both stars have a RV and their probabilities from both the first and second Bayesian analyses.  The projected physical separation was determined by taking the angular separation of each pair in arcseconds and multiplying it by the distance to the primary star in parsecs.  The primary star was determined using the GAIA G magnitude where available, otherwise a V magnitude from SUPERBLINK was used.  The pairs are listed in order of their probabilities from the second distance check.  We note two important details about the catalog.  (1.) Some of the second pass probabilities are zero in the table.  This is due to their distance differences being large (around 500 pc).  (2.) Pairs made of stars from nearby clusters (notably the Pleiades) are part of the table, and can be noticed in Figure \ref{fig:vettedallsky}; no effort was made to remove them.

\subsubsection{Catalog of Unvetted Pairs}
We also present the catalog of the 57,506 unvetted pairs that had first-pass Bayesian probabilities $>10\%$ of being physical pairs, but did not pass the parallax error cut we set in place.  We list all of these pairs in Tables \ref{tab:primaryfailedtab}, \ref{tab:secondaryfailedtab} and \ref{tab:failedbinarytab}.  These tables provide the same information as the tables from section 5.1 with the exception of Table \ref{tab:failedbinarytab} which does not include the second pass Bayesian probability as these pairs do not have this information.

\subsection{Verification of the Wide Binary Status: Radial Velocity Analysis}
\begin{figure}[ht]
\centering
\subfigure{
\includegraphics[scale=0.5]{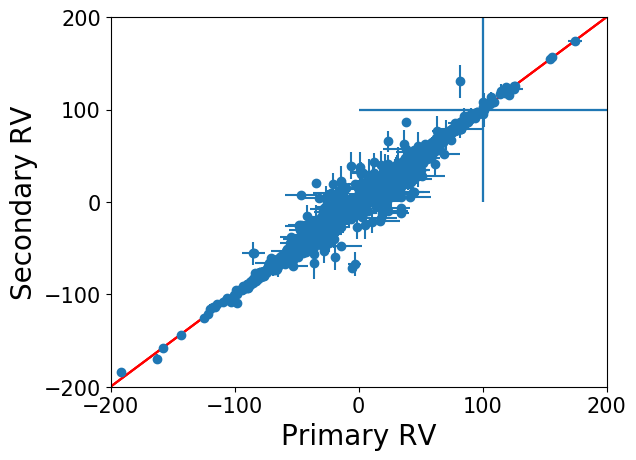}}
\subfigure{
\includegraphics[scale=0.5]{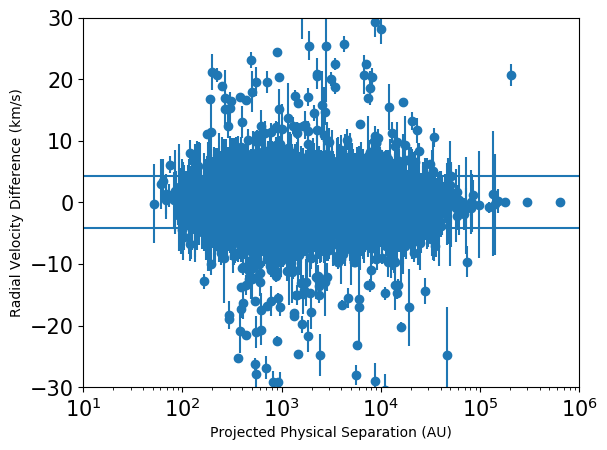}}
\caption{\label{fig:RVdiffyes}Comparison of the radial velocities for the 5,430 wide binaries with probabilities greater than $95\%$ from the second Bayesian analysis and where both components have RVs provided by Gaia DR2.  Top:  Primary RV against Secondary RV.  The red line represents the one to one relation between the two.  If the pairs are binaries, they should be centered around this line which is what we observe.  Bottom:  Radial Velocity differences plotted against the estimated projected separation of the pair.  Our method appears to work well even at higher separations as there are fewer mismatched RVs there. Lines represent the three sigma range for the distribution; 88.2\% of the stars fall within this range.}
\end{figure}

The second Gaia data release contains median radial velocities for around 7 million sources \citep{2019Katzgaiarvs,2018croppergaiaspectrometer,2018sartorettispectro}.  We took our sample of 99,203 pairs with second pass probabilities greater than $95\%$ and searched for pairs with radial velocities in DR2.  This yielded a list of 5,479 pairs for which radial velocities are listed in Gaia DR2 for both components .  If these pairs are true binaries, then one would expect their radial velocities to be similar.  To examine this, we compare the radial velocity of the primary against the radial velocity of the secondary as shown in Figure \ref{fig:RVdiffyes}.  If these are true common proper motion pairs, the points should lie along a straight line.  The top panel of Figure \ref{fig:RVdiffyes} shows that the majority of our points line up as expected.  To examine this more closely, we subtract the radial velocity of the primary and secondary and plot the resulting values as a function of projected physical separation of the pair (Figure \ref{fig:RVdiffyes},bottom panel).  The projected physical separation is calculated in the plane of sky and uses the distance to the primary as the distance to both stars.  This distance is simply multiplied by the angular separation to get the projected physical separation.  Fitting a Gaussian to the distribution yields a sigma of 1.4 $km/s$, which is consistent with the radial velocity errors quoted in Gaia DR2.  Assuming pairs that have radial velocity differences more than 3 sigma are not real binaries, we infer a possible containmination rate of $11.8\%$ for the $>95\%$ group.  This could be over-estimating the containmination rate for several reasons.  For the above estimate, we do not include an error cut for the radial velocities so we could be getting radial velocity differences with large errors.  We tried this same analysis using pairs where the radial velocity error was less than 3 for both pairs and got a lower containmination rate of $5.3\%$.  Another reason is that we may be detecting the orbital motions of some of these pairs.  To examine this possibility, we looked at the radial velocity difference as a function of projected physical separation, shown in bottom panel of Figure \ref{fig:RVdiffyes}.  If orbital motions were significant, we should see more discrepant radial velocity differences at lower separations as the orbital motion should be larger.  We, however, see no such dependence, which suggests that orbital motion has little significance.  Finally, another reason for a large velocity difference may be that one of the components of the wide binary hides as an unresolved spectroscopic sub-system.

\begin{figure}[ht]
\centering
\subfigure{
\includegraphics[scale=0.5]{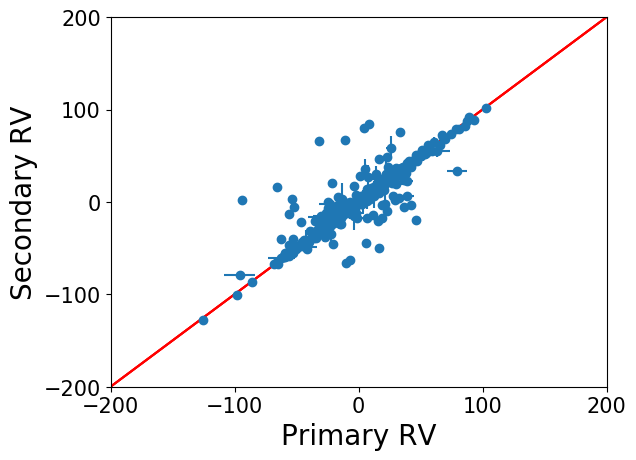}}
\subfigure{
\includegraphics[scale=0.5]{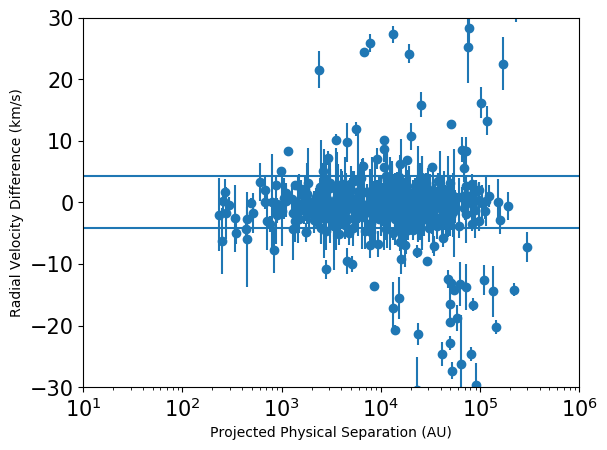}}
\caption{\label{fig:RVdiffmaybe}Comparison of the radial velocities for the 470 wide binary candidates with probabilities between $95\%$ and $20\%$ from the second Bayesian analysis and where both components have RVs provided by Gaia DR2.  Top:  Primary RV against Secondary RV.  The red line represents the one to one relation between the two.  If the pairs are binaries, they should be centered around this line.  Bottom:  Radial Velocity differences plotted against the estimated projected separation of the pair.  Lines represent the three sigma range for the distribution of pairs with probabilities $>95\%$; 77.1\% of the stars fall within this range.}
\end{figure}

\begin{figure}[ht]
\centering
\subfigure{
\includegraphics[scale=0.5]{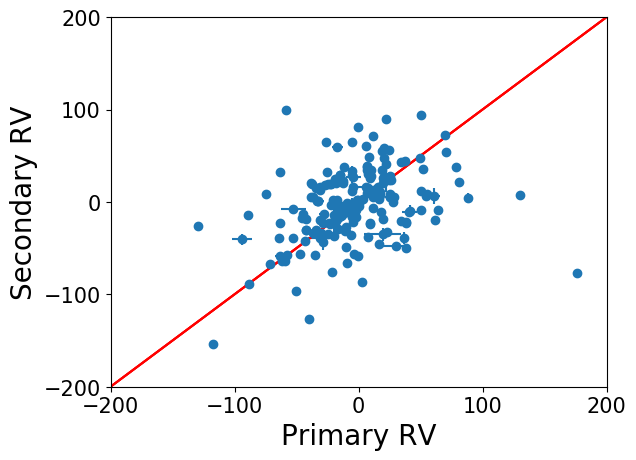}}
\subfigure{
\includegraphics[scale=0.5]{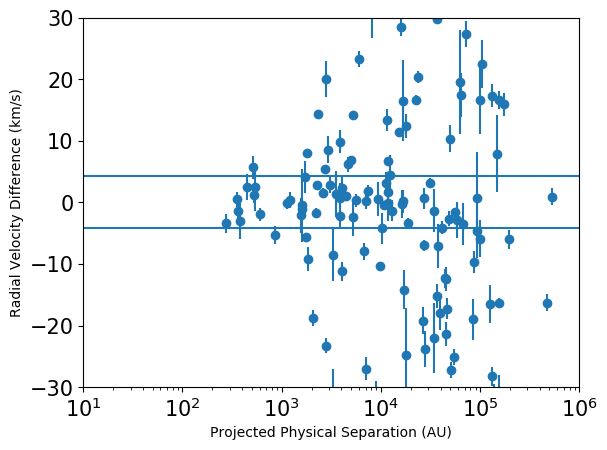}}
\caption{\label{fig:RVdiffno}Comparison of the radial velocities for the 199 wide binary candidates with probabilities between less than 20\% from the second Bayesian analysis and where both components have RVs provided by Gaia DR2.  Top:  Primary RV against Secondary RV.  The red line represents the one to one relation between the two.  If the pairs are binaries, they should be centered around this line.  Bottom:  Radial Velocity differences plotted against the estimated projected separation of the pair.  Lines represent the three sigma range for the distribution of pairs with probabilities $>95\%$; 37.7\% of the stars fall within this range.}
\end{figure}

For comparison, we provide the same plots using the two other probability groups: the pairs with $20\%-95\%$ probability of being real binaries, and the pairs with $< 20$\% probability.  These are shown in Figures \ref{fig:RVdiffmaybe} and \ref{fig:RVdiffno}.  The lower number of pairs with radial velocities in these two groups is because there are less pairs overall in these probability ranges.  Examining Figures \ref{fig:RVdiffmaybe} and \ref{fig:RVdiffno} shows that as the probability decreases, the spread in the radial velocities increases.  This is mirrored in the precentage outside the three sigma lines which for pairs with probabilities between $20\%-95\%$ is $24.8\%$, and for pairs with probabilities $< 20\%$ is $76.9\%$.  The coincidence in radial velocities for many of the pairs suggests that there are still real pairs in these probability ranges.  Once the radial velocities are released for more of the catalog, real pairs can be identified more easily in these low probability regimes.

\begin{figure}[ht]
\centering
\subfigure{
\includegraphics[scale=0.5]{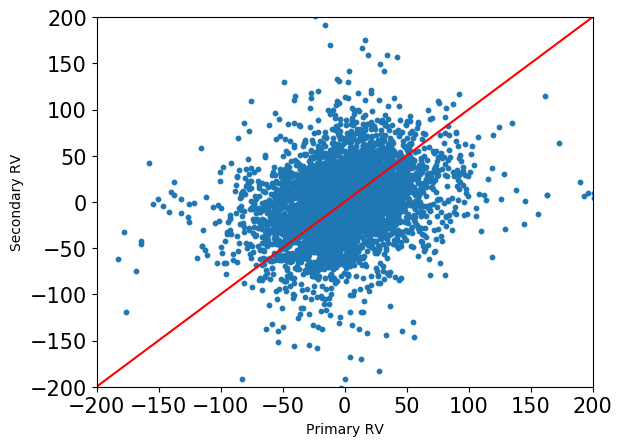}}
\subfigure{
\includegraphics[scale=0.5]{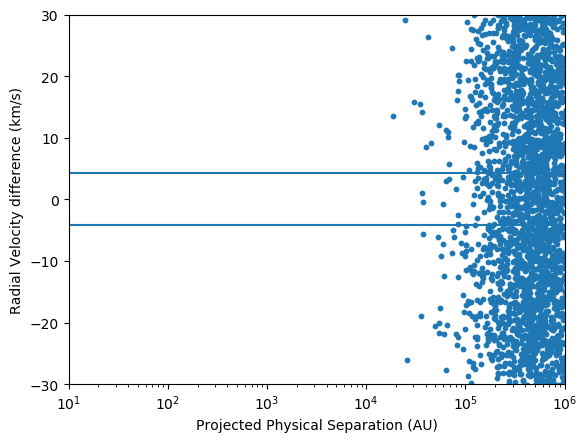}}
\caption{\label{fig:RVdiffbad}Comparison of the radial velocities for 5,000 pairs that have probabilities from the first pass Bayesian analysis of less than $0.5\%$ and both components have RVs from Gaia DR2.  Top:  Primary RV against Secondary RV.  The red line represents the one to one relation between the two.  If the pairs are binaries, they should be centered around this line.  Bottom:  Radial Velocity differences plotted against the estimated projected separation of the pair.  Lines represent the three sigma range for the distribution of pairs with probabilities $>95\%$.}
\end{figure}

To compare this distribution to what one would expect from pure chance alignments of unrelated stars, we select pairs that were rejected in the first pass for having very low ($<< 1$\%) probabilities of being binaries. We examine a subset of 5,000 such pairs for which we found Gaia DR2 for both stars.  The results are shown in Figure \ref{fig:RVdiffbad}.  Most pairs in this group have projected physical separations around $10^6$ AU and are widely distributed in rv difference confirming that they are chance alignments.   Interestingly these chance alignment pairs show a broad correlation in their radial velocities, but with an overall dispersion in RV differences of 8.8 km/s, larger than the Gaia errors. This correlation is clearly not because the pairs are physical binaries, instead we believe that field star radial velocities are broadly correlated with each other in different parts of the sky, in part due to solar reflex motion, and in part due to local stars being organized in stellar streams. The bottom panel of Figure \ref{fig:RVdiffbad} reveals that these pairs have very large separations and must be chance alignments.  

\begin{figure}[ht]
\centering
\subfigure{
\includegraphics[scale=0.5]{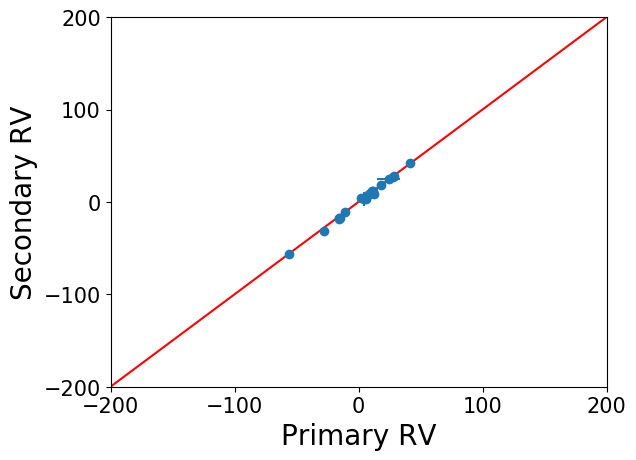}}
\subfigure{
\includegraphics[scale=0.5]{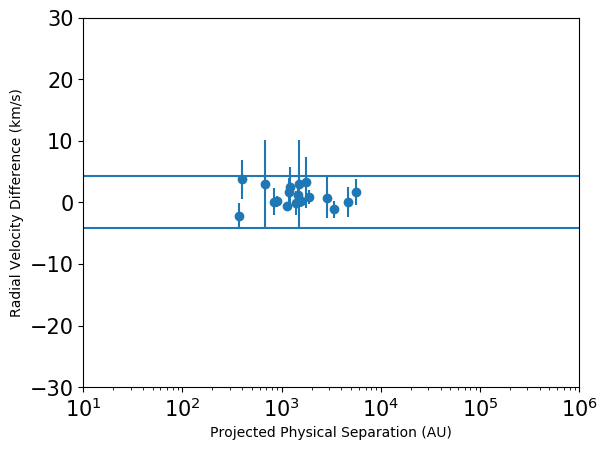}}
\caption{\label{fig:RVdiffunvetted}Comparison of the radial velocities for 19 pairs from the unvetted sample where both components have Gaia RVs.  Top:  Primary RV against Secondary RV.  The red line represents the one to one relation between the two.  If the pairs are binaries, they should be centered around this line.  Bottom:  Radial Velocity differences plotted against the estimated projected separation of the pair.  Lines represent the three sigma range for the distribution of pairs with probabilities $>95\%$.}
\end{figure}

We also include a comparison of radial velocities for pairs in the unvetted subset, for which Gaia radial velocities were also found for a few pairs; this is shown in Figure \ref{fig:RVdiffunvetted}.  This subset includes pairs that had probabilities above 10\% from the first pass of the Bayesian analysis but had high parallax errors or no parallax for one of the components.  Although only 19 of the wide binaries were found to have radial velocities for both components, these 19 all appear to be real pairs real pairs, as demonstrated by the close coincidence in their radial velocities. This increases our confidence that a significant number of stars in the unvettted list are physical binaries as well.  

\subsection{Examining the Sample}
\subsubsection{Projected Physical Separations Analysis}

\begin{figure*}
\centering
\subfigure{
\includegraphics[scale=0.35]{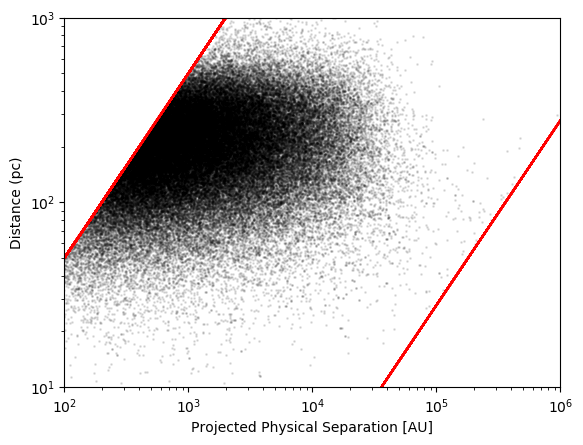}}
\subfigure{
\includegraphics[scale=0.35]{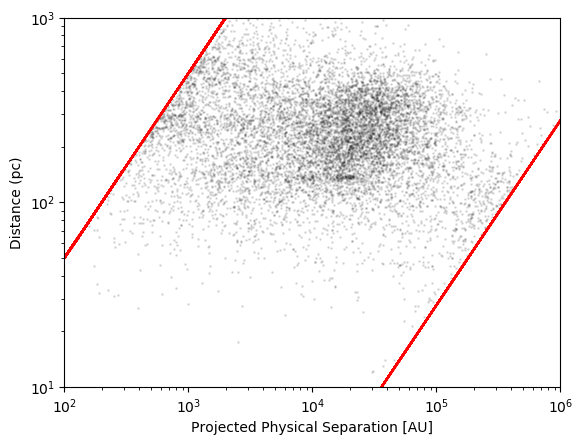}}
\subfigure{
\includegraphics[scale=0.35]{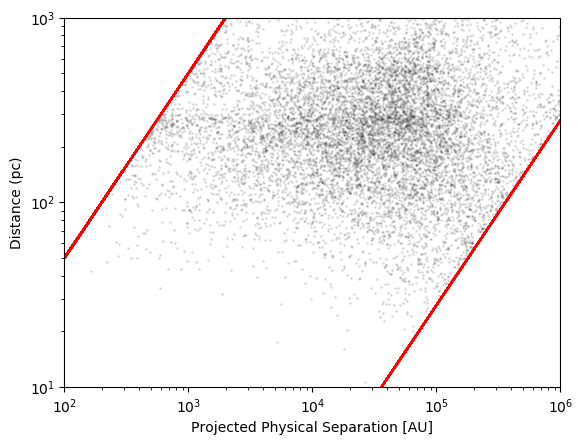}}
\subfigure{
\includegraphics[scale=0.35]{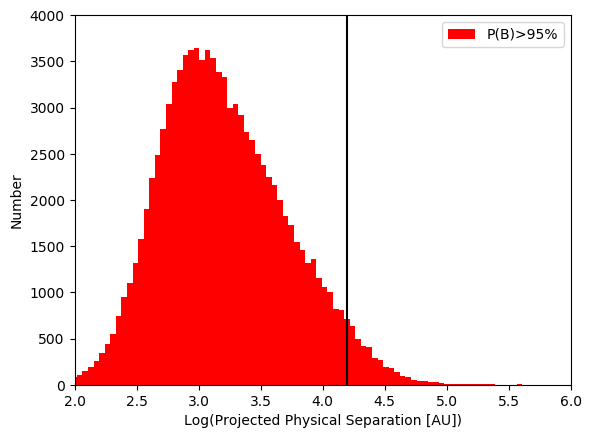}}
\subfigure{
\includegraphics[scale=0.35]{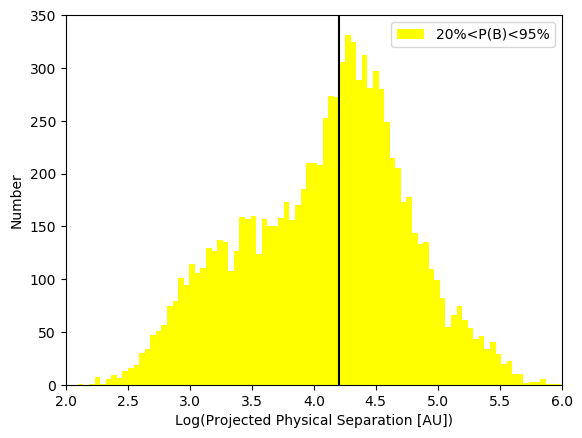}}
\subfigure{
\includegraphics[scale=0.35]{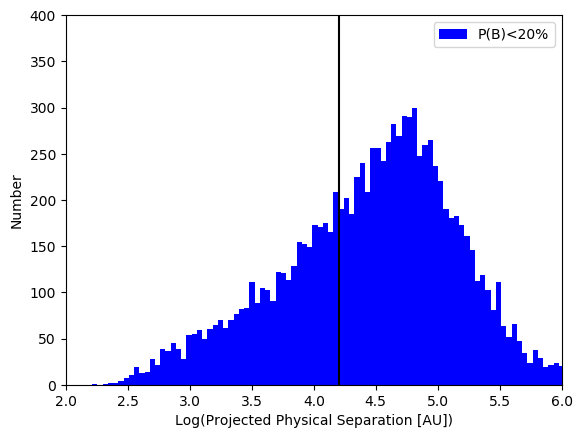}}
\caption{\label{fig:projsepforgroups}Upper row: Projected physical separation vs distance for candidate wide binaries identified in our Bayesian analysis.  The sharp edge on the left is due to our 2" cutoff in angular separation while the diagonal line in the lower right is from our 1 degree search limit; both are marked with a red line. Left: the "Yes" group of pairs with probability $>95\%$ of being real binaries. Middle: the "Maybe" group of pairs with probability $20-95\%$ of being real binaries. Right: the "No" group of pairs with probability $<20\%$ of being real binaries.  Lower row:  Histograms of projected physical separation for the three groups listed above.  As probability decreases from left to right, the peak of the histogram shifts to higher projected physical separations, but the samples are increasingly contaminated by chance alignments based on our analysis.  This means the secondary peak at large separations is likely not real. The black line in the lower plots represent the line of bimodality suggested \citet{2010Dhital}}
\end{figure*}

\begin{figure}
    \centering
    \includegraphics[width=3in]{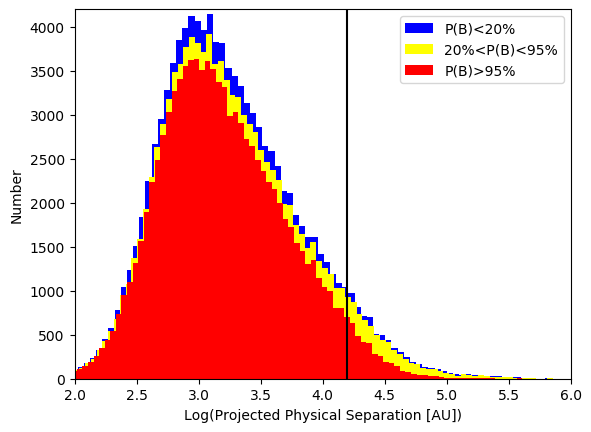}
    \caption{\label{fig:togetherphysepproportional}Combined projected physical separation distribution for all pairs in the SUPERWIDE sample with a weight added which is their probability of being a gravitationally bound system.  The red histogram represents pairs with probabilities > 95\%, the yellow histogram represents the additional pairs with probabilities between 95\% and 20\%, and the blue histogram represents the additional pairs with probabilities < 20\%.  This combined histogram shows no evidence of a secondary peak beyond $10^{4.2}$ AU separation, as suggested in other studies.  The black line represents the line of bimodality suggested \citet{2010Dhital}.}
\end{figure}

Figure \ref{fig:projsepforgroups} shows the histogram of the projected physical separation for pairs with second pass probabilities, i.e. pairs from the parallax vetted subset, and breaks down the physical separations into three groups based on their probability of being wide binaries.  The "Yes" group corresponds to pairs with probabilities greater than $95\%$, the "Maybe" group consists of those pairs with probabilities between $95 - 20\%$ and the "no" group contains pairs with probabilities less than $20\%$.  The upper row of Figure \ref{fig:projsepforgroups} shows the distribution of the three probability groups in plots of distance to the primary vs. projected physical separation.  As seen in the plots, most of the stars in our sample have distances between 100 - 300 pc, largely due to the high proper motion limit of the search catalog.  The red lines show our angular separation limits.  The inner line corresponds to our adopted 2" anuglar separation limit, to account for the fact that Gaia does not completely detect all pairs below that level.  The outer line shows our 1 degree search radius limit.  The bottom row of Figure \ref{fig:projsepforgroups} displays the distribution of projected physical separations for our three probability-selected groups.  The black lines in the plots in the bottom row of Figure \ref{fig:projsepforgroups} represent the line of bimodality suggested by \citet{2010Dhital} at $10^{4.2}$ AU projected physical separation, which hypotheically the tail end of the "normal" wide binary distribution from the population of extremely wide, co-moving pairs.  The left panel shows that the highest probability pairs form a single peak with no sign of having a bimodal distribution.  This is in constrast to previous surveys \citep{2015slowII,2017Oh,2017Oelkers} that suggested a more clearly bimodal distribution, with an increase in the number of wide pairs from $10^4$ AU up to the parsec scale and beyond.  What we find is that this second population of very-wide pairs does not show up in the high probability ("Yes") group. A bimodal distribution does appear to emerge in the lower probability subsets ("Maybe", "No") with a second peak starting to appear at higher separations (around $10^{4.2}$ AU) in the intermediate probability bin (middle panel) and then shifting to larger separations in the lowest probability bin (right panel).   However, one has reasons to doubt whether this shows evidence of a distinct population due to much smaller numbers of pairs in these low-probability subsets.  As a point of fact, Figure \ref{fig:togetherphysepproportional} shows the combined distribution of projected physical separations with a weight added to take into account the probability of the pair.  For example, if a pair has a probability of 20\%, it counts for 0.2 in this figure.  This figure shows that although some pairs are added at large separations, those appear to just be a continuation of the tail end of the distribution of "normal" wide binaries.  This is especially the case for the pairs with probabilities $<20\%$: in Figure \ref{fig:projsepforgroups}, these pairs show a peak at $10^5$ AU, but once the probability weight is added in Figure \ref{fig:togetherphysepproportional}, this peak vanishes showing that most of those pairs had extremely low probabilities of being gravitationally bound systems, i.e. most of them are simply consistent with being chance alignments.

We believe the reason for the apparent bi-modality in Figure \ref{fig:projsepforgroups} is that as the value of our second pass probability decreases, the pairs go from being dominated by genuine gravitationally bound wide binary candidates to being increasingly contaminated by chance alignments, which can have, or appear to have, parsec-scale separations.  The peak in the distribution continues to shift to larger separations from the "maybe" to the "no" probability groups because it is a mix of the continuing tail end of the distribution and chance alignments.  In the "no" group, the majority of these pairs are chance alignments, which is why in Figure \ref{fig:togetherphysepproportional} when probability is added as a weight, the combined distribution of projected physical separations appears as a single distribution with most of the contribution occurring at $10^3$ AU.  If a second population was involved, we would expect the contribution from the lowest probability bin to be focused at larger separations rather than what is seen in Figure \ref{fig:togetherphysepproportional}.
Based on Figure \ref{fig:togetherphysepproportional}, it appears that the true wide binary distribution consists of a single peak, which is largely determined by the lower detection limit on angular separation.  Because of the absence of a second peak in our "Yes" group, we are confident that (1.) there is no secondary population of extremely-wide, parsec-scale, gravitationally bound pairs, and (2.) our survey identifies most of the real, gravitationally bound binaries. This, combined with the confirmation of binaries from our radial velocity analysis, makes use confident that our "Yes" sample constitutes a "clean" sample of wide binaries, with minimal contamination from chance alignments. We will note two potential biases in this analysis.  (1.) In the design of our two part analysis, we took only the pairs that had first pass probabilities $>10\%$ for the second pass.  It is possible that some of the roughly 556,900,000 possible pairs not included in the second pass could have ended up with second pass probabilities between 50\% and 10\%.  These could contribute additional pairs to the tail of the distribution but they would be low probability pairs.  (2.) Our sample is based on a catalog of high proper motion stars, most of which should not be young stars.  Young stars would make up the majority of the co-moving pairs described in \citet{2017Oh} as they are cluster members and pairs that could be the remnants of wide binaries.  More analysis on this is planned.

\subsection{Over-luminous Components in Wide Binaries: The "Lobster" Diagram}

\begin{figure*}
\centering
\subfigure{
\includegraphics[scale=0.5]{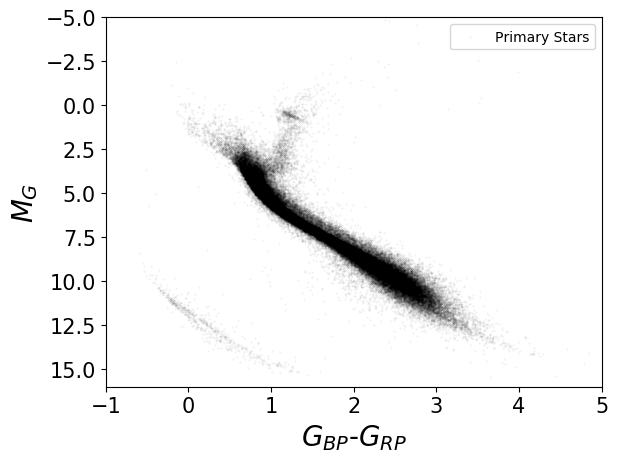}}
\subfigure{
\includegraphics[scale=0.5]{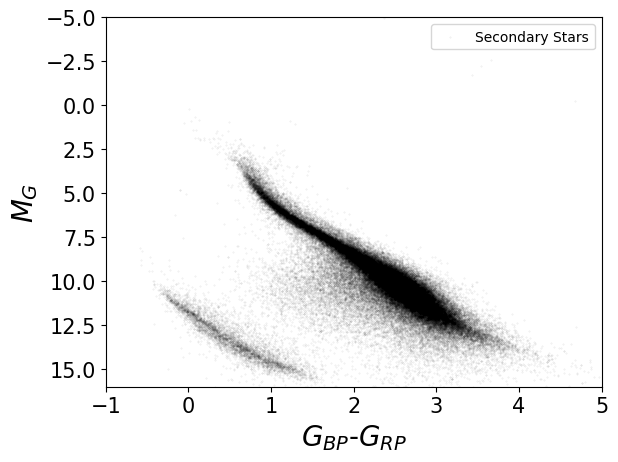}}
\caption{\label{fig:priandsechr}Color-magnitude diagrams for the pairs in our "Yes" subset, i.e. pairs with Bayesian probabilities $>95\%$ of being wide physical binaries.  Left: Color-magnitude diagram for the primary components.  Right: Color-magnitude diagram for the secondary components. Primary stars are found of all types, including notable subsets of red giants, subgiants, more massive main-sequence stars, and white dwarfs. Secondaries are overwhelmingly low-mass stars and white dwarfs.}
\end{figure*}

With the accurate parallaxes provided by Gaia DR2, we are able to examine the color-magnitude diagrams of the components (primary and secondary) of our candidate wide binaries in detail.  Figure \ref{fig:priandsechr} shows the separate color-magnitude diagrams for the primaries and secondaries of our "Yes" group, i.e. the subset with Bayesian probabilities $>95\%$ of being true binaries.  We do require that the both components in each pair have a $G_{BP}$ and $G_{RP}$ magnitude from Gaia for this analysis.  This requirement eliminates some pairs that have components identified from the SUPERBLINK catalog but are not listed in Gaia DR2.  The main sequence in both cases is well defined, however, the color-magnitude diagram of the secondaries does suggest that our subset may be including "unclean" stars from Gaia, by that we mean stars found in between the main sequence and the white dwarf cooling sequence, a zone that is not expected to contain any significant number of stars.  We believe there are several possible explanations for this unwanted component.  One is that these are chance alignments and represent unrelated background stars.  The parallax of the secondary could, in this case, be wrong and simply matches the parallax of the primary.  Another explanation is that these are true secondaries whose Gaia parallaxes are incorrect.  However, we do not believe this to be likely as subbing the primaries parallax for the secondaries keeps these stars in the same location in the color-magnitude diagram.  A third possibility is that the $G_{BP} - G_{RP}$ colors of the secondaries are incorrect, specifically that they are bluer than the actual values.  Assuming the Gaia parallaxes and magnitudes are accurate, however, the a fourth and most likely explanation is that these secondaries are in fact unresolved pairs of white dwarf + M dwarfs that have blended colors. 

\begin{figure}
\centering
\includegraphics[width=3.3in]{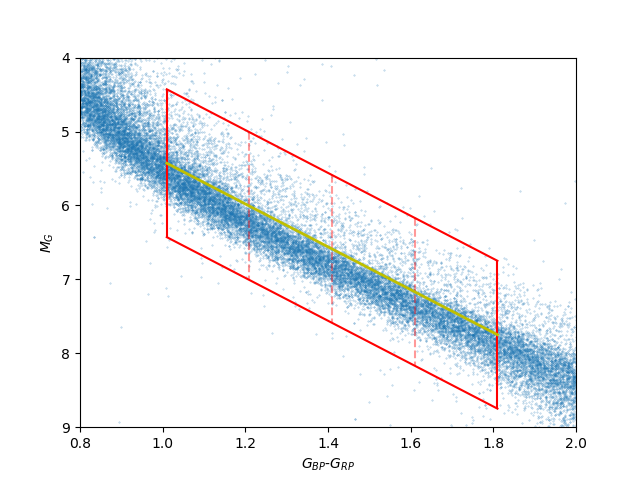}
\caption{\label{fig:overlumhr}Color-magnitude diagram for primary stars focusing on the K dwarf region of the main sequence, which shows a simple, near-linear color-magnitude relationship.  The red box shows the region being examined while the red dotted lines show the regions which we use to analyze the change in unresolved binary fraction along the K dwarf sequence.  The yellow line through the middle of the sequence represents our arbitrary reference line used to calculate the "over-luminosity factor" of every component in the wide binaries. The broadening of the main-sequence due to metallicity variations ("cosmic scatter") and the dedoubling of the main sequence due to unresolved components (luminosity booster) are both noticeable on the diagram. The "overluminosity factor" is a combination of both effects.}
\end{figure}

In both panels of Figure \ref{fig:priandsechr} in the color range of $G_{BP} - G_{RP} \sim$ 1.5, one notices a doubling of the main sequence with an $\sim$0.7 magnitude upward shift, consistent with the presence of additional companions that are not resolved by Gaia.  The same effect was also noted by \citet{2018elbadry} in their own catalog.  To investigate this interesting feature, we examine a sample of 2,227 K+K wide binaries with primary distances less than 250 pc and and Bayesian probabilities $>99$\% of being physical binaries; we use this more restrictive subset to minimize contamination from chance alignments.  On the assumption that some of the objects may be unresolved systems, we modify our defininition of ``primary'' and ``secondary'' by using color instead of magnitude and defining the bluer star to be the primary component.  Figure \ref{fig:overlumhr} zooms in on the K dwarf locus (red box) on the color-magnitude diagram for these high probability wide binaries.  The K dwarf color range was provided by the Leonardo Paredes.  They used a sample of vetted single stars within the 25 pc RECONS sample with known spectral types classified by \citet{2009gray}.  They obtained astrometry and photometry from Gaia DR2 for those stars and then matched the spectral types to different colors and absolute magnitudes to define the K dwarf limits.  A problem with the identification of over-luminous objects (due to unresolved companions) is the magnitude of the "cosmic scatter", which is due to metallicity differences between the local field stars and which significantly broadens the main sequence in particular for M dwarfs, but also in the K dwarf regime. To disentangle both effects (metallicity and multiplicity), we use the following procedure. First, we define an "over-luminosity factor" ($F_{OL}$)
\begin{equation*}
F_{OL} = M_G - [M_G]_{Kref}
\end{equation*}
which which is the difference between the absolute magnitude $M_G$ of a star and a reference level $[ M_G ]_{Kref}$ meant to represent the color-magnitude relationship for single-star K dwarfs of an arbitrary metal abundance. For this we adopt the relationship:
\begin{equation*}
[ M_G ]_{Kref} = 2.9 ( G_{BP} - G_{RP} ) + 2.5
\end{equation*}
This relationship is represented by the yellow line in Figure \ref{fig:overlumhr}.  This line roughly represents the division between the single star main sequence and the unresolved binary main sequence, although this choice is arbitrary.

\begin{figure}
\centering
\includegraphics[width=3.3in]{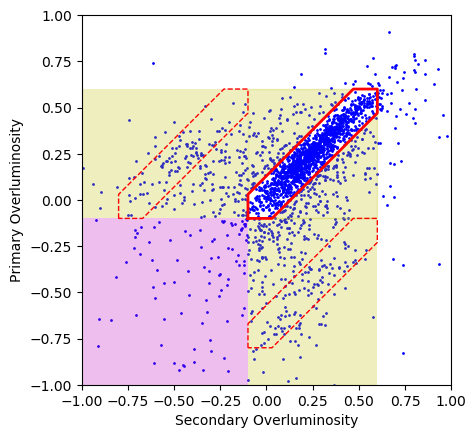}
\caption{\label{fig:lobster}The "Lobster Diagram" showing the over-luminosity of the primary component plotted against the over-luminosity of the secondary component, in high-probability wide binary systems.  Because the components of wide binaries have similar metallicity, their over-luminosity factors are strongly correlated (if both components are single) and they fall on a 1-1 sequence, the "body" of the lobster.  The purple shaded region represents the area of the plot where both components are over-luminous and hence the wide binaries is potentially a quadruple system.  The yellow shaded regions represent areas where either one of the components is an unresolved binary, meaning the systems is, in fact, a triple.  The two red bordered regions in the yellow shaded regions represents the area where equal mass unresolved binaries exist.}
\end{figure}

Figure \ref{fig:lobster} shows the distribution of the over-luminousity factor $F_{OL}$ of the primaries as a function of the $F_{OL}$ of the secondaries.  The red bordered region going from roughly 0,0 to 1,1 represents components of the wide binaries that are ``single''.  The correlation between the $F_{OL}$ values of the primaries and secondaries here represents the effect of the "cosmic scatter": stars of low metallicity in our subset have $F_{OL}\sim0.6$, while stars of high metallicity have $F_{OL}\sim-0.1$.  This range explains the $\sim0.7$ magnitude spread of the single star main sequence.  Wide systems whose components are single stars cluster along this line because the metallicity of both components are the same, and thus the overluminosity of the primary correlates with the overluminosity of the secondary.  The yellow shaded regions represents areas where one of the components appears to be over-luminous as compared to its companion, and thus likely is a triple system.  The area to the left of the single-star locus on the diagram is where the secondary is over-luminous while while the area below the single-star locus is where the primary is over-luminous.  The purple shaded region on the lower left is where one would expect a pair to be if both components are over-luminous and the wide binary is actually a quadruple system.  The red bordered regions inside the yellow-shaded regions on the diagram represent areas where one would expect a pair to be if one of the component is an unresolved, equal-mass binary, i.e. two stars of the same luminosity.  If these are equal mass systems, then the orbital separation is expected to be small, making them excellent targets for future spectroscopic binary surveys. The pairs located between the single-star locus and equal-mass binary loci are likely unresolved binaries of unequal mass, and would also make excellent targets for binary star searches in general.

With this method, we can determine that of the subset of 2,227 ``extremely likely'' K+K wide systems, 1,343 show no evidence of either component having an unresolved companion and thus are likely to be mostly ``true'' binaries, i.e. systems of only two widely separated individual stars.  On the other hand, we find that 449 are systems with an over-luminous primary star while 339 are systems with an over-luminous secondary.  In addition, we find 96 systems showing signs of being quadruple systems (both components over-luminous).  These numbers suggest that the higher order multiplicity fraction of our K+K wide binaries is at least 39.6\%.  We stress that this is most likely an underestimate.  There will be high delta-mag companions that will not contribute enough light to be picked up by Gaia.  In addition, we know of pairs with angular separations between 2-10 arcseconds that have a third companion at a larger separation and these are not accounted here.  On the other hand, some factors could also cause a star to appear over-luminous while not being an unresolved binary.  These include a star evolving off the main sequence, a pre-main sequence star still in the contraction phase, or errors in the Gaia measurements.  For the first two alternatives, we believe that such cases should not be happening in this particular subset because our survey is using a proper motion limited sample, which reduces the number of young stars, and we are focusing on the K dwarfs which should not be evolving off the main sequence yet.  We believe the third problem is mitigated by the parallax error cut that we implemented before the second pass of the analysis.

\section{Discussion}

\subsection{Comparison to Previous Searches}
With the growing availability of large catalogs, there has been a renewed interest in examining these catalogs for wide binaries that appear as common proper motion companions.  There are two big differences between our analysis and these previous searches.  The first is that we are focusing our search on a proper motion limited subset, whereas previous searches looked at all stars in a target catalog.  Focusing on stars with proper motions greater than 40 mas/yr makes it easier to pick out wide pairs due to a reduced amount of contamination from distant field stars.  The second distinction is that our Bayesian analysis uses an empirical approach to determine the probability distributions of binaries and chance alignments, which is lifted out of the data. This contrasts to other approaches that attempt to model these distributions using theoretical or semi-empirical considerations. Our analysis also does more than just using simple cuts in proper motion and separation space, and instead assigns probabilities for all pairs over a broader search range. 

\begin{figure}[ht]
\centering
\includegraphics[width=3in]{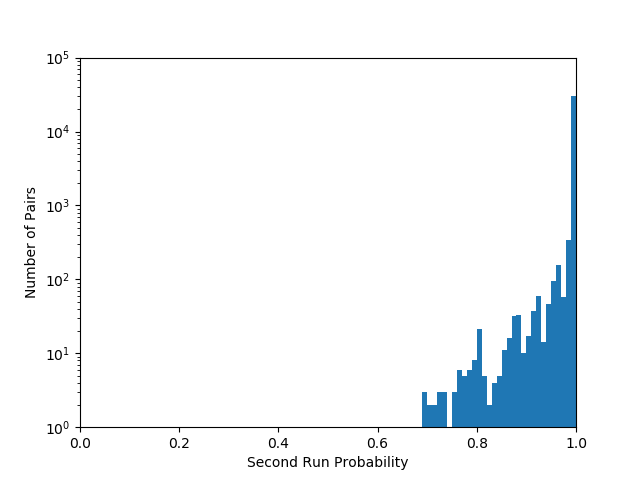}
\caption{\label{fig:elbadryprobs}Second pass Bayesian probabilities for wide binaries in SUPERWIDE and are also in \citet{2018elbadry}.  Almost all have probabilities $>95\%$.}
\end{figure}

\begin{figure*}
\centering
\subfigure{
\includegraphics[scale=0.5]{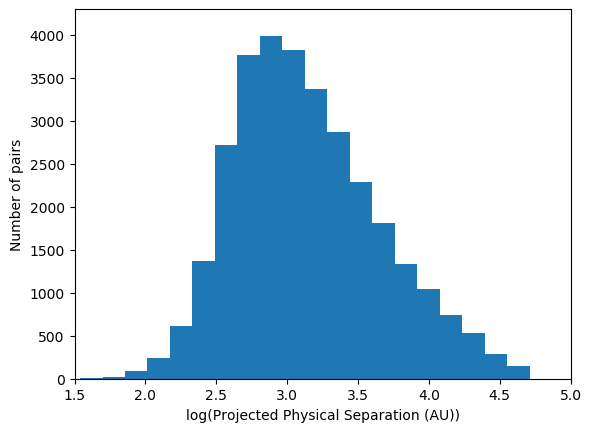}}
\subfigure{
\includegraphics[scale=0.5]{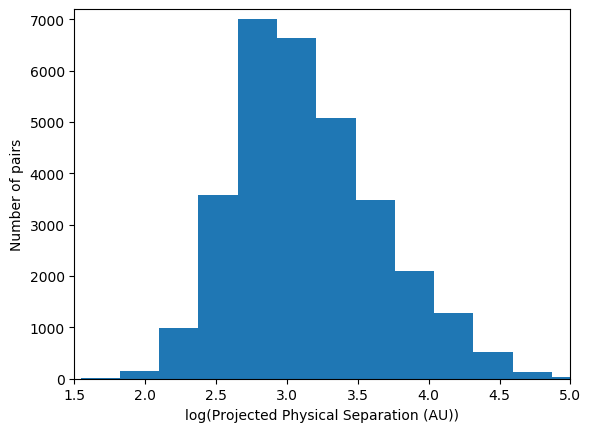}}
\caption{\label{fig:physepelbadry}Projected physical separation of our wide binaries candidates that were previously identified in the \citet{2018elbadry} (left panel) and our larger subset of candidates that were not identified by \citet{2018elbadry} (right panel).  The distribution in the right panel extends farther in project physical separation range compared with \citet{2018elbadry} which has a sharp cut at 50,000 AU.  }
\end{figure*}

For example, the catalog of \cite{2018elbadry} contains 55,128 binaries from the Gaia DR2 catalog.  31,536 of those pairs have proper motions above 40 mas/yr for both stars in the pair.  We take these pairs and match them against the 119,316 pairs that made our parallax control cut and were run through the second pass of the Bayesian analysis. We find 31,066 pairs in common between the two sets, which shows that our methods recover essentially all the \citet{2018elbadry} pairs.  Of the remaining 470 unmatched pairs, most are not found in our catalog because they either did not pass the requirement that the parallax error be less than $10\%$ of the parallax itself or they fell below the 2" limit we set for our pairs.  Figure \ref{fig:elbadryprobs} shows the histogram of probabilities that we assigned to each of the 31,066 pairs in common between the two sets.  As seen, the vast majority of the pairs are found to have high probabilities of being binaries in our second pass analysis. This suggests that the \citet{2018elbadry} analysis identifies the most obvious pairs, but fails to recover substantial numbers of potential systems.  Figure \ref{fig:physepelbadry} compares the projected physical separation histograms for the subset of our wide binaries that are in \citet{2018elbadry} (left panel) and for the subset of wide binaries that are not in \citet{2018elbadry} (right panel).  The two plots look nearly identical with the only difference being that our lower probability sample extends to larger physical separations as one might expect as we do not include a physical separation cut. Both distributions still peak around $10^3$ AU however, and both have an exponential decay at higher separations.    

\subsection{Higher Order Multiplicity of K+K Wide Binaries}
The higher order multiplicity fraction for the widest K+K systems has potential implications for determining how these wide systems formed.  The unfolding of triple systems scenario \citep{2012Reipurth} consists of three stars forming in a single protostellar cloud.  Over time, two of the stars form a close binary and kick the third out into a higher orbit to conserve angular momentum.  If this scenario is the dominant formation mechanism for wide binaries, it is expected that a large fraction of wide binaries should be in triple systems.  Many of the other scenarios predict a large higher order multiplicity fraction as well, \citep{2017Tokovinin,2010Kouwenhoven}.  However, our higher order multiplicity fraction is $39.6\%$.  However, our wide binaries span a wide range of projected physical separation and the pairs with separations $\sim1,000$ AU most likely formed through other methods, i.e. turbulent fragmentation, and not the unfolding of triple systems.  To examine this, we need to examine the widest systems.  Therefore, we took a sample of the K+K wide binaries which had projected physical separations $>10,000$ AU and reran our analysis on this subsample of 112 K+K wide binaries that had these large separations.  We find that 69 are true wide binaries, 23 are possible triples with an over-luminous primary, 16 are possible triples with an over-luminous secondary and 4 are possible quadruple systems.  From these values, the higher order multiplicity fraction of this subset is 38.3\%, essentially the same as the complete sample.  As the separations for these systems are on the order of a typical protostellar core ($\sim0.1$ pc, $\sim20,000$ AU), this seems to suggest that the higher order multiplicity fraction for these K dwarfs is lower than predicted.  However, this requires further follow-up as the over-luminosity factor may be able to find extremely close third companions well($\sim0.1$ arcsecond), but there may not be enough light contributed from potential unresolved companions at larger separations ($\sim0.5$ arcsecond) in Gaia DR2.

In addition, we examined the overall binary fraction of our components, as if all wide primaries and secondaries were independent systems of their own.  This was done by breaking down each pair into its components and calculating the fraction of components that are over-luminous.  For our sample sample of 4,454 individual components in the 2,227 K+K wide systems, we find the overall binary fraction, based on overluminosity, to be 22\%.  For the 112 widest systems, with physical separations >10,000AU, the multiplicity fraction for the 224 components is comparable, at 21\%.  For comparison, solar type stars in the field have a multiplicity fraction of 46\% \citep{2010Raghavan}, while for M dwarfs, the multiplicity fraction is estimated to be 26.8\% \citep{2019wintersmdwarfmult}.  Assuming that the multiplicity fraction for K dwarfs is between these two fractions and that these wide binaries formed widely separated but near each other (adjacent cores scenario, \citet{2017Tokovinin}), it appears that our binary fraction is significantly lower than expected, which would suggest that wide binaries are more often composed of pairs of single stars than one might expect if the components were drawn from the field population.  As discussed above, there are various reasons that our binary fraction could be underestimated. This, once again, points to the need for follow-up observations to look for close companions to these wide binaries.

This analysis has been using K+K wide binaries from across the K dwarf region; because the "primaries" are significantly bluer (and of higher mass) than the "secondaries", the binarity fraction of the primaries may be higher simply because of a mass-dependence on the binary fraction. To investigate this, we break down the K dwarfs to examine how the higher order multiplicity changes as a function of color.  We do this by splitting the K dwarf region into four color bins as seen by the yellow dotted lines in Figure \ref{fig:overlumhr}.  Our four color regions span from 1.01 to 1.21, 1.21 to 1.41, 1.41 to 1.61 and 1.61 to 1.81 in $G_{BP} - G_{RP}$.  In each region, we examine all components (primaries and secondaries) that fall within that range, and calculate the binary fraction from the over-luminosity factor.  Table \ref{tab:homfk} shows our results from this analysis.  We find that the unresolved binary fraction decreases as a function of the component's color/mass, from $\sim$30\% to $\sim19\%$.  This provides further evidence that the binary fraction is generally a function of mass in low-mass stars, with higher mass objects being more likely to be unresolved systems, mirroring what is found for single K dwarfs in the field population.

\section{Summary}

We have searched the high-proper stars in the Gaia catalog to identify 99,203 common proper motion pairs with probability $>95\%$ of being wide binary systems, based on a Bayesian analysis method.  Of those pairs, we estimate that about 364 are expected to be false positives.  The analysis uses a two-step process: a "first-pass" analysis determined the probability of the pair to be a wide binary based on proper motion and angular separation alone, while a "second-pass" analysis compares the parallaxes of the two components, for pairs selected in the first pass.  We present a complete list of the $P>95\%$ systems, along with two other subsets(1) a list of 20,187 candidates with second-pass Bayesian probability $0\%<P<95\%$ of being wide binaries, and (2) a list of 57,506 "unvetted" common proper motion pairs, with high probability of being wide binaries from the first-pass Bayesian analysis, but that could not be verified in the second pass due to missing or uncertain parallax data.  While there are undoubtedly real wide binaries in each of the latter subsets, we caution to users to be careful in using them as the two subsets are likely contaminated by chance alignments.  To verify this, we have checked our catalog using the radial velocities provided by Gaia DR2 to ensure that our catalog consisted of genuine wide binaries.  The spread of radial velocities increases with decreasing Bayesian probability and is consistent with what is expected. In addition, we compared our catalog with an earlier catalog of Gaia wide binaries assembled by \citet{2018elbadry} and find our catalogs to be in agreement.  Most pairs in common between the sample we find to have high probabilities of being genuine pairs with a few exceptions.  

An examination of the projected physical separations of our pairs and our pairs finds that in or best subset of pairs with probability $P>95\%$ (the "Yes" sample), there is no evidence of bi-modality due to a second population of wide binaries with extremely large ($\rho$>100,000AU) projected physical separations, as had been suggested.  Instead, we demonstrate that this hypothesized population of extremely wide systems represents the tail of the ``normal'' wide binary projected physical separation distribution.  

Our investigation into the doubling of the main sequence in the K dwarf region of the color-magnitude diagram reveals that $39.6\%$ of the wide binaries in that region are higher order multiples.  Our over-luminosity factor analysis further reveals that for the widest binaries ($\rho > 10,000$ AU) the higher order multiplicity is $38.3\%$.  This is much lower than predictions on wide binary formation expect, however, we believe that our value is underestimated by a variety of factors.   In addition, we find that further evidence that binary fraction changes with primary color/mass.  

\acknowledgments
We would like to thank the anonymous referee for the their thoughtful comments.  We would like to thank Leonardo Paredes and the RECONS team for the K dwarf color range used to find the K+K wide binaries.  We would like to thank Gerard van Belle, Todd Henry, Elliott Horch, Bokyoung Kim, Maxwell Moe, Ilija Meden, Leonardo Paredes, Andrei Tokovinin, and Erika Wagoner for their insightful  discussions.

This work has made use of data from the European Space Agency (ESA) mission {\it Gaia} (\url{https://www.cosmos.esa.int/gaia}), processed by the {\it Gaia} Data Processing and Analysis Consortium (DPAC, \url{https://www.cosmos.esa.int/web/gaia/dpac/consortium}). Funding for the DPAC has been provided by national institutions, in particular the institutions participating in the {\it Gaia} Multilateral Agreement.

This publication makes use of data products from the Two Micron All Sky Survey, which is a joint project of the University of Massachusetts and the Infrared Processing and Analysis Center/California Institute of Technology, funded by the National Aeronautics and Space Administration and the National Science Foundation.

This material is based upon work supported by the National Science Foundation under Grant No. AST 09-08406. This project was also supported by NASA as part of the K2 Guest Observer program, under grants NNX16AI63G and 80NSSC18K0140.  In addition, this project was supported by the NN-Explore program through RSA 1623647.

\software{Python,Numpy,LMFIT,Scipy,Matplotlib}

\bibliography{sample63}{}
\bibliographystyle{aasjournal}

\begin{deluxetable}{llllllllll}
\tablewidth{700pt}
\tabletypesize{\scriptsize}
\rotate
\tablehead{
\colhead{Catalog ID} & \colhead{Gaia DR2 ID}&
\colhead{RA} & \colhead{Dec} & 
\colhead{$PM_{RA}$} & \colhead{$PM_{Dec}$} & 
\colhead{Parallax} & \colhead{Gaia G} &
\colhead{B-R} & \colhead{Gaia RV}
\\ 
\colhead{} & \colhead{} & \colhead{degrees} & \colhead{degrees} & \colhead{$mas/yr$} & \colhead{$mas/yr$} & 
\colhead{mas} & \colhead{mag} & \colhead{mag} & \colhead{km/s}
} 
\tablecaption{\label{tab:primarytab}Data on primary stars in pairs that have probabilities greater than $10\%$ from the first pass probabilities and have parallax errors less than $10\%$ of the parallax.  We provide the Catalog ID, Gaia ID, RA, Dec, $PM_{RA}$, $PM_{Dec}$, Parallax, Gaia G, Gaia B - Gaia R, Gaia radial velocities for those that have it.  If the component is from SUPERBLINK, Gaia id is set to 99999.99. Pairs are in a 1-1 match with Table \ref{tab:secondarytab} and \ref{tab:binarytab}.  The rest are available online.  If Gaia RV is not present, value is set at 99999.99 in online data table.}
\startdata
SWB1&2132602965008510080&289.40943&49.20537&80.324&112.491&10.9543&14.8427&2.2534&99999.99\\
SWB2&1260355683405766656&212.89005&28.03704&-26.992&46.888&9.4085&8.7861&0.6569&3.18\\
SWB3&6048314340854256640&250.37519&-22.30216&-40.648&-91.051&7.2029&13.4135&2.0112&-14.04\\
SWB4&2543280552367099904&9.94537&0.26607&75.983&-204.794&29.5832&12.4899&2.4828&99999.99\\
SWB5&5787222832248879104&182.16387&-80.4334&-94.52&7.946&15.5051&13.9574&2.4804&99999.99\\
SWB6&2153399712050263808&285.00734&56.96101&-118.31&-162.49&11.2896&8.4766&0.8137&-4.74\\
SWB7&903348277956806528&125.33579&34.30949&-115.451&-111.495&22.3057&8.3081&0.8719&33.09\\
SWB8&2938406277905135232&94.43241&-20.80665&-26.874&65.134&6.668&14.1492&1.8887&99999.99\\
SWB9&2318637789803820800&6.09962&-29.6631&80.394&-135.454&22.7456&12.3974&2.1211&41.36\\
SWB10&151650076838458112&69.20213&27.13156&232.873&-148.136&57.1046&7.7178&1.3751&41.63\\
\enddata
\end{deluxetable}

\begin{deluxetable}{llllllllll}
\tablewidth{700pt}
\tabletypesize{\scriptsize}
\rotate
\tablehead{
\colhead{Catalog ID} & \colhead{Gaia DR2 ID} &
\colhead{RA} & \colhead{Dec} & 
\colhead{$PM_{RA}$} & \colhead{$PM_{Dec}$} & 
\colhead{Parallax} & \colhead{Gaia G} &
\colhead{B-R} & \colhead{Gaia RV}
\\ 
\colhead{} & \colhead{} & \colhead{degrees} & \colhead{degrees} & \colhead{$mas/yr$} & \colhead{$mas/yr$} & 
\colhead{mas} & \colhead{mag} & \colhead{mag} & \colhead{km/s}
} 
\tablecaption{\label{tab:secondarytab}Data on secondary stars in pairs that have probabilities greater than 10\% from the first pass probabilities and have parallax errors less than $10\%$ of the parallax.  We provide the Catalog ID, Gaia ID, RA, Dec, $PM_{RA}$, $PM_{Dec}$, Parallax, Gaia G, Gaia B - Gaia R, Gaia radial velocities for those that have it.  If the component is from SUPERBLINK, Gaia id is set to 99999.99.  Pairs are in a 1-1 match with Table \ref{tab:secondarytab} and \ref{tab:binarytab}.  The rest are available online.  If Gaia RV is not present, value is set at 99999.99 in online data table.}
\startdata
SWB1&2132602965008510592&289.41098&49.20583&88.813&109.051&10.9684&15.3907&2.4927&99999.99\\
SWB2&1260355679110038912&212.89&28.03639&-29.432&44.766&9.4189&9.4469&0.7225&3.52\\
SWB3&6048314340854256512&250.37855&-22.3011&-41.762&-91.413&7.209&15.8155&3.0359&99999.99\\
SWB4&2543281175138179712&9.95544&0.28565&76.514&-205.958&29.6859&14.928&3.2094&99999.99\\
SWB5&5787222832248879488&182.1699&-80.43307&-94.978&10.089&15.5334&16.387&3.1465&99999.99\\
SWB6&2153399712048655104&285.00385&56.96129&-117.512&-161.575&11.3046&13.6938&2.0843&99999.99\\
SWB7&903348273663336448&125.33691&34.31055&-118.379&-106.303&22.364&12.3785&2.3538&99999.99\\
SWB8&2938406277905135488&94.43125&-20.80593&-26.718&64.536&6.6732&17.005&2.4665&99999.99\\
SWB9&2318637785507972736&6.09998&-29.66243&81.437&-131.951&22.8066&12.9529&2.258&41.09\\
SWB10&151650935831913216&69.18817&27.16367&227.512&-148.428&57.4881&15.6252&0.9061&99999.99\\
\enddata
\end{deluxetable}

\begin{deluxetable}{lcccccc}
\tablewidth{500pt}
\tabletypesize{\scriptsize}
\rotate
\tablehead{
\colhead{Catalog ID} & \colhead{Angular Separation} & 
\colhead{Projected Physical Separation} & \colhead{$\Delta G$} & \colhead{RV Difference} &
\colhead{First Run Bayesian Probability} & \colhead{Second Run Bayesian Probability}
\\
\colhead{} & \colhead{"} & \colhead{AU} & \colhead{mag} & \colhead{km/s} & \colhead{\%} & \colhead{\%}
}
\tablecaption{\label{tab:binarytab}Data on binary pairs that have probabilities greater than 10\% from the first pass probabilities and have parallax errors less than $10\%$ of the parallax.  We present the Catalog ID, angular separation, projected physical separation, $\Delta G$ magnitude difference, RV differences where both stars have Gaia RVs, first pass Bayesian probability and second pass Bayesian probability.  If one or both stars do not have RVs, the value is set to 99999.99.  We provide both probabilities to show the effect of adding parallax data to the analysis.  Full table available online.}
\startdata
SWB1&4.01696&366.7&-0.548&99999.99&99.977&99.995\\
SWB2&2.34466&249.21&-0.661&-0.34&99.989&99.995\\
SWB3&11.83587&1643.21&-2.402&99999.99&99.979&99.995\\
SWB4&79.27159&2679.62&-2.438&99999.99&99.999&99.995\\
SWB5&3.79452&244.73&-2.43&99999.99&99.996&99.995\\
SWB6&6.92542&613.43&-5.217&99999.99&100.0&99.995\\
SWB7&5.09035&228.21&-4.07&99999.99&99.998&99.995\\
SWB8&4.67795&701.55&-2.856&99999.99&99.994&99.995\\
SWB9&2.66177&117.02&-0.556&0.27&100.0&99.995\\
SWB10&123.9735&2170.99&-7.907&99999.99&99.997&99.995\\
\enddata
\end{deluxetable}

\begin{deluxetable}{llllllllllllllll}
\tablewidth{700pt}
\tabletypesize{\scriptsize}
\rotate
\tablehead{
\colhead{Gaia DR2 ID} & \colhead{RA} & \colhead{Dec} & 
\colhead{$PM_{RA}$} & \colhead{$PM_{Dec}$} & 
\colhead{Parallax} & \colhead{Gaia G} &
\colhead{B-R} & \colhead{Gaia RV}
\\ 
\colhead{} & \colhead{degrees} & \colhead{degrees} & \colhead{$"/yr$} & \colhead{$"/yr$} & 
\colhead{"} & \colhead{mag} & \colhead{mag} & \colhead{km/s}
} 
\tablecaption{\label{tab:primaryfailedtab}Data on primary stars in pairs that have probabilities greater than 10\% from the first pass probabilities but have parallax errors $>10\%$ of the parallax itself.  We provide the Gaia DR2 ID, RA, Dec, $PM_{RA}$, $PM_{Dec}$, Parallax, Gaia G, Gaia B - Gaia R, Gaia Radial Velocities for those that have it.  If the component is from SUPERBLINK, Gaia id is set to 99999.99.  Pairs are in a 1-1 match with Table \ref{tab:secondaryfailedtab} and \ref{tab:failedbinarytab}.  The rest are available online.  If Gaia RV is not present, value is set at 99999.99 in online data table.}
\startdata
5884478552748243584 & 235.93267 & -54.90283 & 14.949 & 37.923 & 4.5261 & 13.89 & 1.576 & 99999.99\\
5833123388322264064 & 239.86951 & -60.05021 & 6.271 & 44.897 & 7.0731 & 10.546 & 0.991 & 3.28\\
4453039448459006848 & 242.91703 & 9.25112 & -4.736 & 45.532 & 4.6534 & 18.595 & 2.928 & 99999.99\\
6244478004203627264 & 244.90327 & -20.62286 & 7.687 & 48.412 & 5.0736 & 17.01 & 2.998 & 99999.99\\
5930816954967570944 & 250.33166 & -52.91031 & 26.907 & 30.751 & 6.1025 & 18.506 & 1.358 & 99999.99\\
5927411041532098048 & 251.9549 & -57.94281 & 12.719 & 39.242 & 4.8924 & 10.435 & 0.839 & 83.45\\
5802320913607276672 & 253.868 & -74.11293 & -4.084 & 42.243 & 2.2223 & 15.096 & 1.292 & 99999.99\\
4050723913317960064 & 272.26664 & -28.82923 & 1.765 & 40.333 & 3.9614 & 16.134 & 1.993 & 99999.99\\
4279542045514285056 & 282.5737 & 3.43835 & 4.723 & 40.162 & 2.4528 & 16.737 & 1.983 & 99999.99\\
6435293368119930112 & 288.88332 & -65.21554 & -29.785 & 41.396 & 7.0742 & 17.07 & 3.055 & 99999.99\\
6639697317069560832 & 289.42898 & -57.09519 & -21.928 & 36.862 & 5.2488 & 13.312 & 1.579 & 64.37\\
\enddata
\end{deluxetable}

\begin{deluxetable}{lllllllll}
\tablewidth{700pt}
\tabletypesize{\scriptsize}
\rotate
\tablehead{
\colhead{Gaia DR2 ID} & \colhead{RA} & \colhead{Dec} & 
\colhead{$PM_{RA}$} & \colhead{$PM_{Dec}$} & 
\colhead{Parallax} & \colhead{Gaia G} &
\colhead{B-R} & \colhead{Gaia RV}
\\ 
\colhead{} & \colhead{degrees} & \colhead{degrees} & \colhead{$"/yr$} & \colhead{$"/yr$} & \colhead{"} & \colhead{mag} & \colhead{mag} & \colhead{km/s}
} 
\tablecaption{\label{tab:secondaryfailedtab}Data on secondary stars in pairs that have probabilities greater than 10\% from the first pass probabilities but have parallax errors $>10\%$ of the parallax itself.  We provide the Gaia DR2 ID, RA, Dec, $PM_{RA}$, $PM_{Dec}$, Parallax, Gaia G, Gaia B - Gaia R, Gaia Radial Velocities for those that have it.  If the component is from SUPERBLINK, Gaia id is set to 99999.99.  Pairs are in a 1-1 match with Table \ref{tab:secondarytab} and \ref{tab:binarytab}.  The rest are available online.  If Gaia RV is not present, value is set at 99999.99 in online data table.}
\startdata
5884477075272635136 & 235.93088 & -54.90432 & 16.564 & 40.291 & 2.836 & 20.491 & 1.975 & 99999.99\\
5833125552965965696 & 239.64202 & -60.06028 & 4.819 & 46.269 & 7.5865 & 19.0 & 1.3 & 99999.99\\
4453039448459006976 & 242.91631 & 9.25154 & -3.829 & 46.018 & 4.952 & 20.197 & 2.798 & 99999.99\\
6244477999906665088 & 244.90201 & -20.62171 & 5.307 & 48.244 & 3.8703 & 20.518 & 2.258 & 99999.99\\
5930817126751717504 & 250.30199 & -52.91493 & 33.559 & 33.945 & 15.5122 & 19.976 & 0.0 & 99999.99\\
5927411045842815232 & 251.95393 & -57.94086 & 13.588 & 38.302 & 4.3049 & 16.996 & 0.0 & 99999.99\\
5802320810523986304 & 253.86236 & -74.1142 & -4.399 & 40.621 & 2.2216 & 18.811 & 2.059 & 99999.99\\
4050723814577442304 & 272.26422 & -28.83028 & 3.01 & 42.074 & 3.4632 & 18.149 & 2.04 & 99999.99\\
4279541667555198720 & 282.56844 & 3.41689 & 4.482 & 42.817 & 2.8045 & 19.97 & 3.024 & 99999.99\\
6435293363822847232 & 288.8572 & -65.21587 & -30.272 & 41.572 & 8.1237 & 20.446 & 0.936 & 99999.99\\
6639697385790611328 & 289.4246 & -57.08382 & -20.192 & 34.984 & 7.3508 & 20.382 & 0.375 & 99999.99\\
\enddata
\end{deluxetable}

\begin{deluxetable}{lllll}
\tablewidth{700pt}
\tabletypesize{\scriptsize}
\rotate
\tablehead{
\colhead{Angular Separation} & \colhead{Projected Physical Separation} & \colhead{$\Delta G$} & \colhead{RV Difference} & \colhead{First Run Bayesian Probability}\\
\colhead{"} & \colhead{AU} & \colhead{mag} & \colhead{km/s} & \colhead{\%}
}
\tablecaption{\label{tab:failedbinarytab}Data on unvetted pairs.  We present the angular separation, projected physical separation, G magnitude difference, RV difference and first run Bayesian probability.  If one or both stars do not have RVs, the value is set to 99999.99.  Full table available online.}
\startdata
6.52032 &1440.6 &-6.601 & 99999.99 & 99.964\\
410.3998 &58022.62 &-8.453 & 99999.99 & 18.494\\
3.00066 & 644.83 & -1.602 & 99999.99 & 99.999\\
5.91019 & 1164.89 & -3.508 & 99999.99 & 99.988\\
66.51647 & 10899.87 & -1.469 & 99999.99 & 36.155\\
7.26125 & 1484.19 & -6.561 & 99999.99 & 99.994\\
7.20618 & 3242.67 & -3.715 & 99999.99 & 99.992\\
8.51248 & 2148.86 & -2.015 & 99999.99 & 99.976\\
79.53044 & 32424.35 & -3.233 & 99999.99 & 89.143\\
39.43964 & 5575.14 & -3.376 & 99999.99 & 99.669\\
\enddata
\end{deluxetable}

\begin{deluxetable}{cccc}
\tabletypesize{\scriptsize}
\tablewidth{700pt}
\rotate
\tablehead{
\colhead{Region 1} & \colhead{Region 2} & \colhead{Region 3}& \colhead{Region 4}\\
\colhead{1.01<$G_{BP} - G_{RP}$<1.21} & \colhead{1.21<$G_{BP} - G_{RP}$<1.41} & \colhead{1.41<$G_{BP} - G_{RP}$<1.61}& \colhead{1.61<$G_{BP} - G_{RP}$<1.81}
}
\tablecaption{\label{tab:homfk}Higher order multiplicity fraction as a function of K dwarf color}
\startdata
30.4\% & 18.2\% & 17.8\% & 19.1\%\\
\enddata
\end{deluxetable}
\end{document}